\documentclass[aps,prx,reprint,10pt,groupedaddress,showpacs,twocolumn]{revtex4-1}
\usepackage[latin1]{inputenc}
\usepackage[pdftex]{graphicx}
\usepackage[cmex10]{amsmath}
\usepackage{amssymb}
\usepackage{pifont}
\usepackage{psfrag}
\usepackage{amssymb}
\usepackage{amscd}
\usepackage{mathtools}
\usepackage{eucal}
\usepackage{color}
\usepackage{bm}
\usepackage{hyperref}

\DeclareMathOperator{\diag}{diag}

\begin{document}

\newcommand{\J}{{\mathbb J}}
\newcommand{\bi}{\bibitem}
\newcommand{\Kf}{K\hspace{-0.8mm}f}

\title{Finite-time Correlations Boost Large Voltage-Angle Fluctuations in Electric Power Grids}
\author{Melvyn Tyloo\textsuperscript{1,2}, Jason Hindes\textsuperscript{3}, and Philippe Jacquod\textsuperscript{2,4}}
\affiliation{$^{1}$Theoretical Division, Los Alamos National Laboratory, Los Alamos, NM 87545 USA.\\
$^2$Department of Quantum Matter Physics, University of Geneva, CH-1211 Geneva, Switzerland\\
$^3$ U.S. Naval Research Laboratory, Washington, DC 20375, USA \\
$^4$School of Engineering, University of Applied Sciences of Western Switzerland HES-SO CH-1951 Sion, Switzerland }

\date{\today}

\begin{abstract}
Decarbonization in the energy sector has been accompanied by an increased penetration of new renewable energy sources 
in electric power systems.
Such sources differ from traditional productions in that, first, they induce larger, undispatchable fluctuations
in power generation and second, they lack inertia. Therefore, substituting new renewables for traditional generation induces stronger and more
frequent disturbances and modifies the way disturbances
propagate across AC electric power grids. Recent measurements have indeed reported long,
non-Gaussian tails in the distribution of local grid-frequency data. 
Large frequency deviations may induce grid instabilities, leading in worst-case scenarios to 
cascading failures and large-scale blackouts. In this manuscript, we investigate
how correlated noise disturbances, characterized by the cumulants of their distribution, propagate through meshed, high-voltage power grids. We show that 
for a single source of fluctuations, non-Gaussianities in the form of finite skewness and positive kurtosis of the noise distribution propagate over the entire network when the noise correlation time is larger than the 
network's intrinsic time scales, but that they vanish over short distances if the noise fluctuates rapidly. 
We furthermore show that a Berry-Esseen theorem
leads to the vanishing of non-Gaussianities as the number of uncorrelated noise sources increases.  Our results 
show that the persistence of non-Gaussian fluctuations of 
feed-in power have a global impact on power-grid dynamics when they fluctuate over time scales larger than the intrinsic time scales
of the system, which, we argue, is the relevant regime in real power grids. Our predictions are corroborated by 
numerical simulations on realistic models of power grids. \end{abstract}

\maketitle

\section{Introduction}
The fight against climate change is one of the biggest challenges currently facing humankind~\cite{IPCC19}. 
Globally increasing atmospheric and oceanic temperatures have been directly related to the emission of
greenhouse gases~\cite{Fle98}. Therefore, key to mitigating climate changes is our ability to reduce
emissions of such gases. Of particular interest is carbon dioxide, because of its large emission volumes and
century-long lifetime in the atmosphere. 
Decarbonization, i.e. the reduction of carbon dioxide emissions from human activities, 
requires a fast, fundamental shift to renewable, low-carbon energy sources and a systematic electrification of the energy sector.
This will affect the operation of AC electric power grids as both productions and consumptions will change~\cite{AnnualEnOutlook}.
In particular, higher penetration of
renewable energy sources means fluctuating and uncertain power productions~\cite{Mil13}, as well as
reduced electromechanical inertia~\cite{Mac08,Ulb14}, which impacts  
dynamic properties of power systems. It is expected -- and in fact already observed -- that future power grids
will be subjected more often to stronger external perturbations
to which they may respond more strongly~\cite{Aue17,Scha17}.

AC power grids are technological entities that can be modeled as network-coupled dynamical systems. 
Since they operate according to market rules, it is often difficult to obtain true, reliable data on their operational state. Only recently has it been
possible to get access to sufficiently large, statistically significant frequency datasets. Analyses of these datasets
have emphasized the non-Gaussian nature of frequency fluctuations in AC power grids~\cite{Hae18,Ryd20,Ryd21,Sch18},
with distributions exhibiting long tails and large increments. The source of these large deviations
is often attributed to the presence of new renewable sources of energy~\cite{Mil13,Hae18,Wol19}. Large frequency deviations
are an important risk factor for the stability and hence for the safety of operation of present and future AC electric power systems.
It is therefore of the utmost importance to understand how non-Gaussian disturbances propagate through electric networks. 
Many recent papers have investigated
the propagation of disturbances originating from noisy power feed-in into power grids~\cite{Sia16,Ket16,Hae18,Pag19,Tyl19,Tum19,Wol19,Sch20}.
However, most of them considered Gaussian noise, with two notable exceptions. First,
Ref.~\cite{Hae18} showed analytically that the variance of noise-induced
frequency fluctuations decays exponentially away from its feed-in source, while  the frequency kurtosis was numerically observed 
to exhibit a slower, possibly power-law decay. Second, 
Ref.~\cite{Wol19} conjectured that the structure of power grids amplifies non-Gaussianities in power feed-in. 
Different noise probability distributions have been considered, yet the influence of fundamental noise characteristics such as correlation time, or the presence of multiple, independent noise sources has been neglected so far. Below we show that
these characteristics are indeed key to understanding how non-Gaussian fluctuations of voltage angles propagate through
high-voltage power grids.

 \begin{figure*}[t!]
\includegraphics[width=0.19\textwidth]{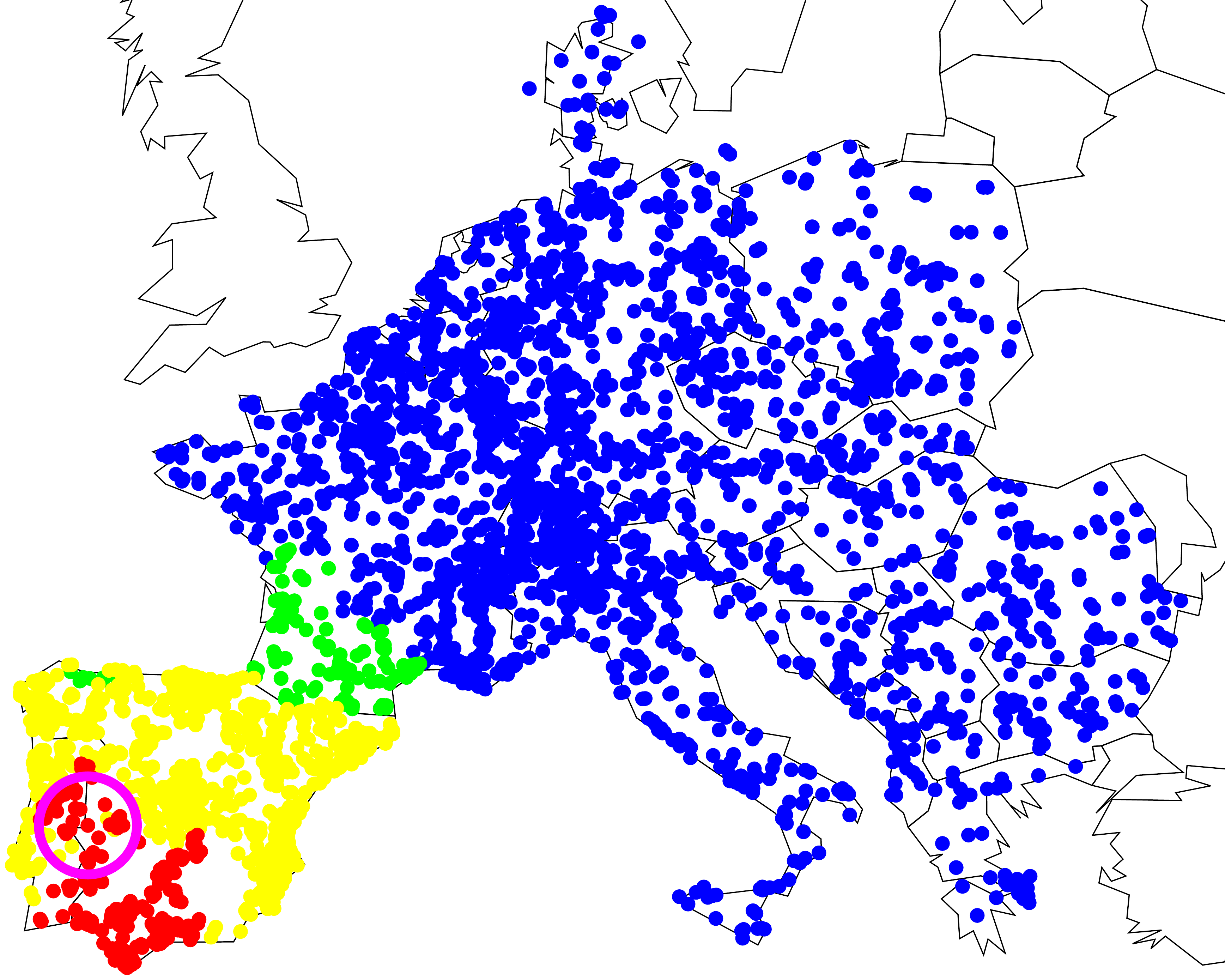}
\includegraphics[width=0.19\textwidth]{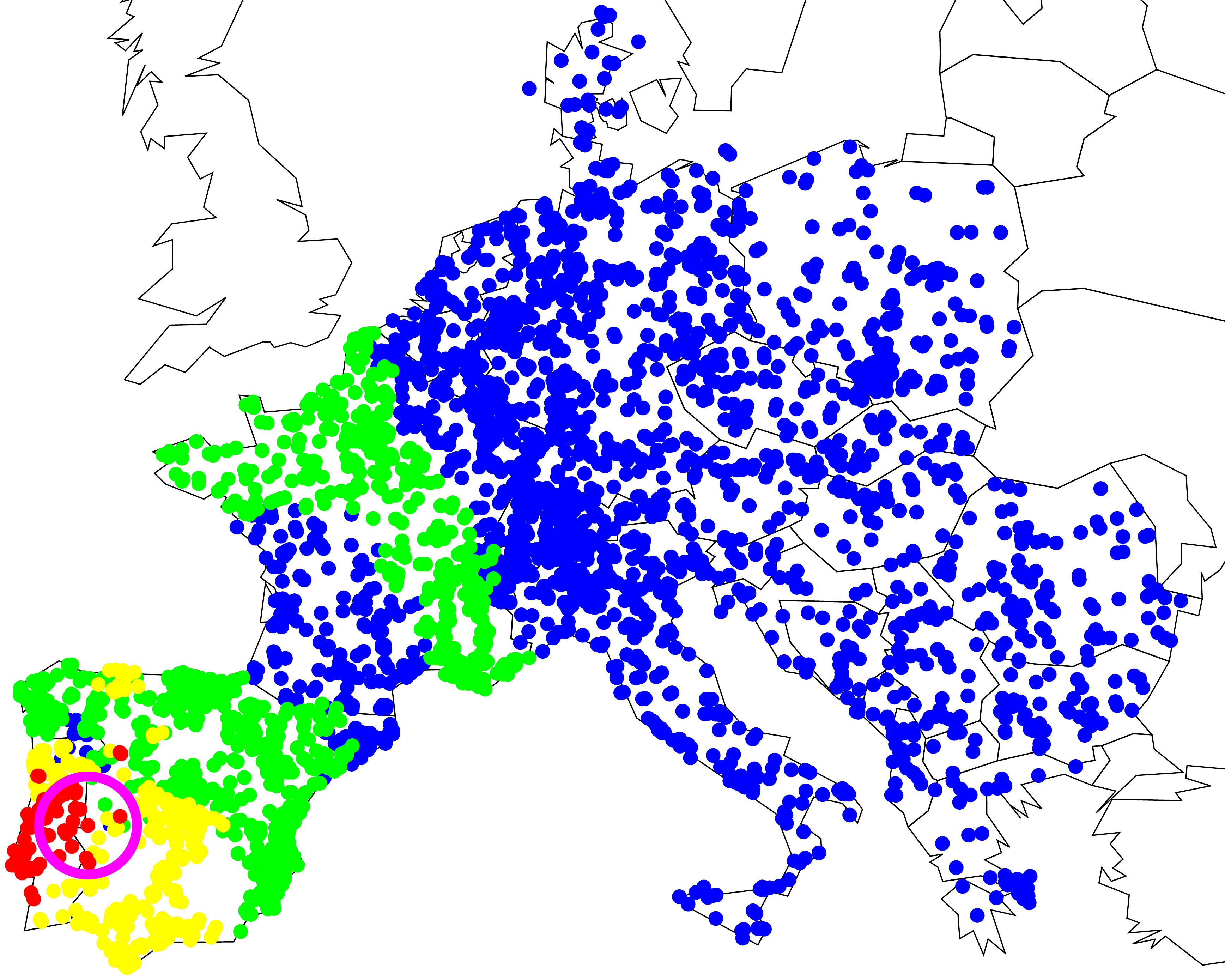}
\includegraphics[width=0.19\textwidth]{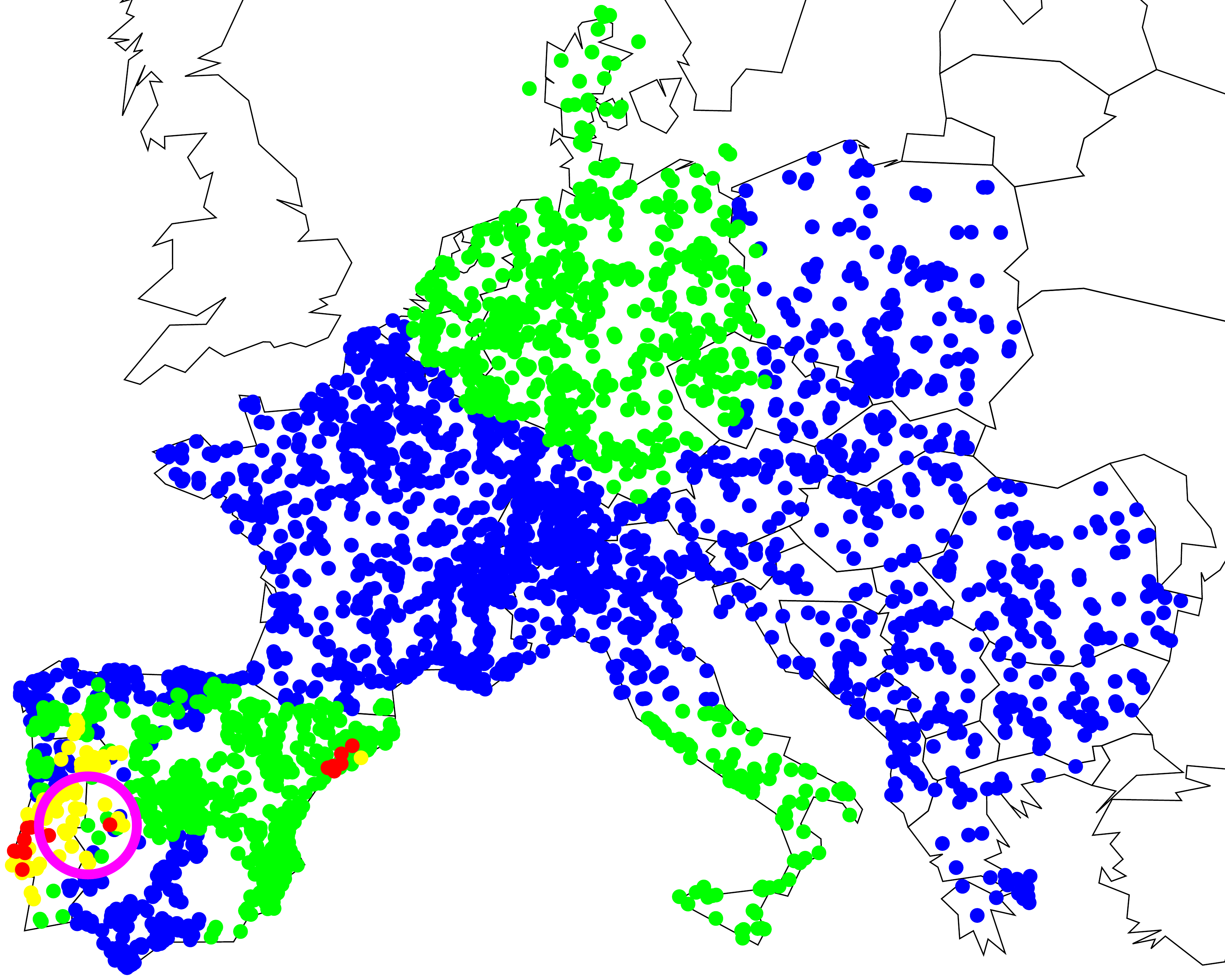}
\includegraphics[width=0.19\textwidth]{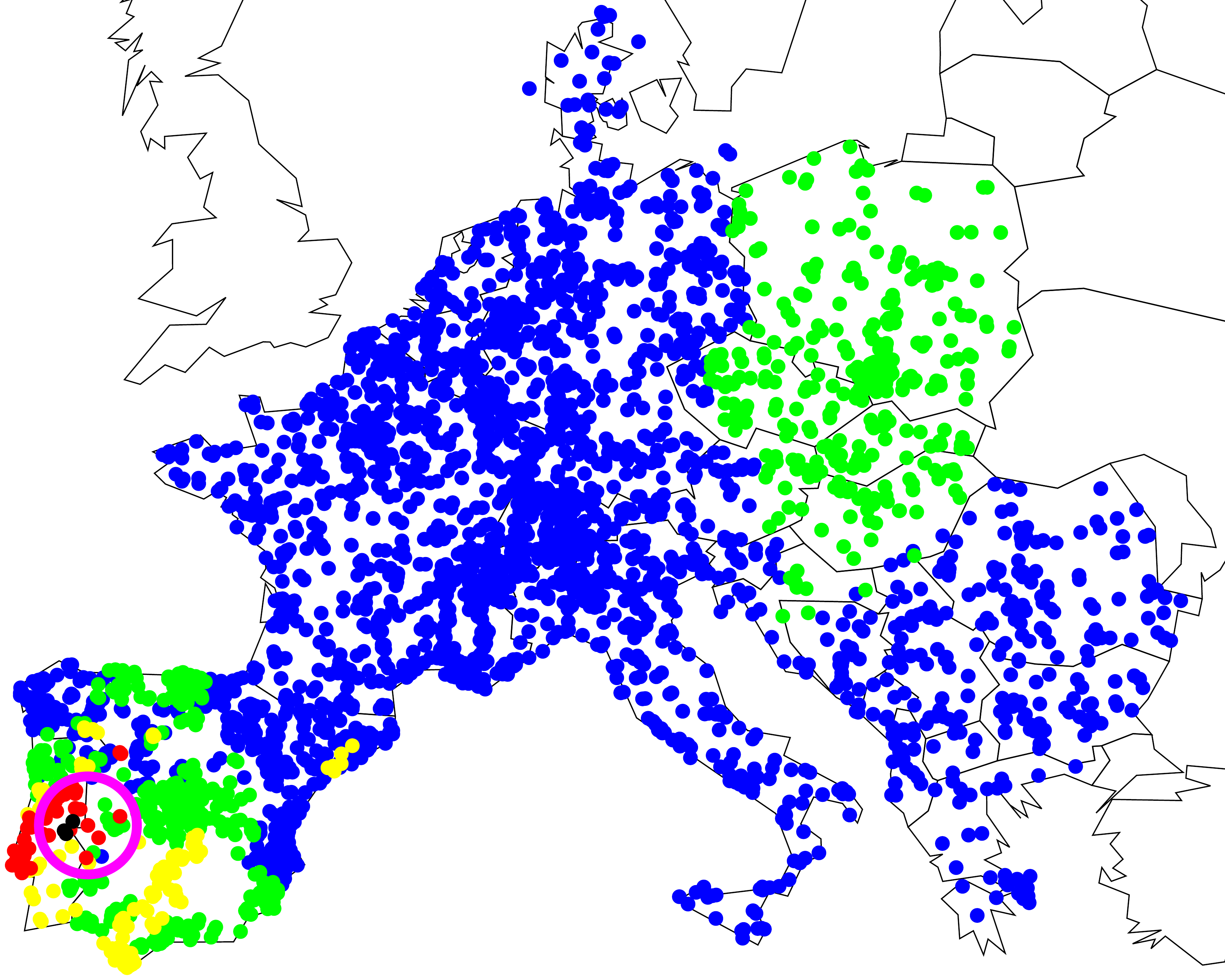}
\includegraphics[width=0.19\textwidth]{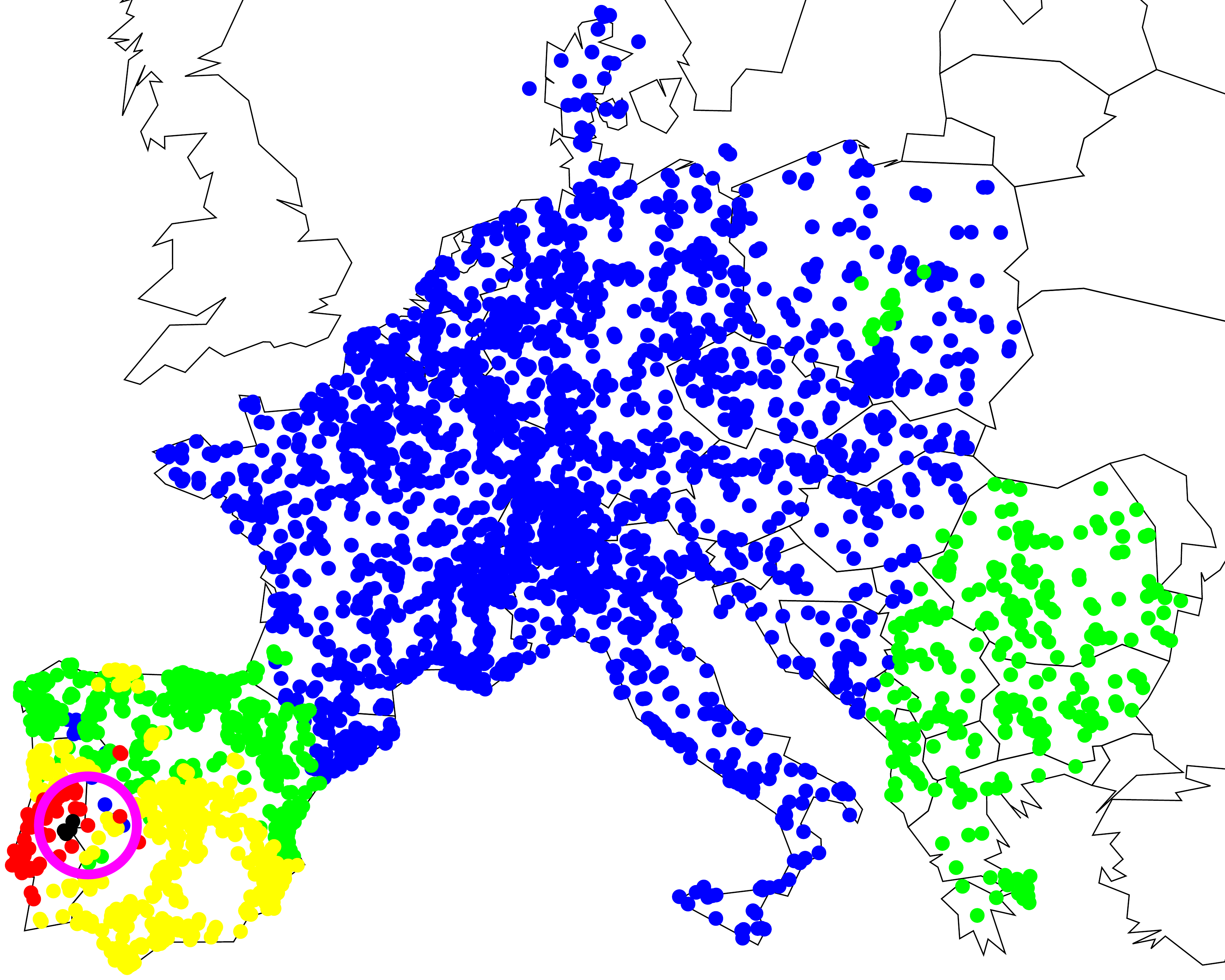}\\
\includegraphics[width=0.75\textwidth]{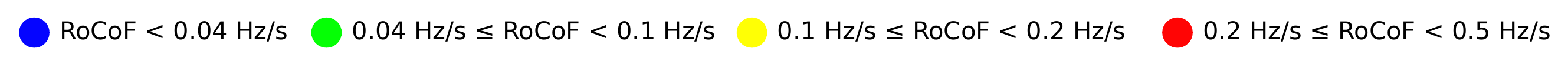}
\caption{Calculated spatio-temporal evolution of the voltage angle disturbance across the 
synchronous grid of continental Europe. Color-plotted
are the local Rates of Change of Frequency (RoCoF, defined on the $i^{\rm th}$ node as $\partial^2_t \theta_i(t)$) for an abrupt power loss of $\delta P=900$ MW in the Iberian Peninsula (location indicated by the
pink circle). Panels correspond to 
snapshots giving the maximal RoCoF over time intervals 0-0.5[s], 0.5-1[s], 1-1.5[s], 1.5-2[s] and 2-2.5[s] from left to right.}
\label{fig:RoCoF_snapshots}
\end{figure*}

The short-time dynamics of power grids is commonly 
modelled by the swing equations~\cite{Mac08}, which are nonlinear, damped wave equations on discrete networks. 
Source terms, representing fluctuating power feed-in, 
generate voltage angle and frequency waves that spread through the system. 
In this manuscript we investigate the propagation of such waves through realistic power grids
and characterize 
the noisy source terms by the cumulants of their distributions and by their correlation time $\tau_0$. 
Given a single, or several sources of non-Gaussian noise, we calculate cumulants $\langle \delta \theta_i^p \rangle_c$, $p \le4$ of the 
voltage angle distributions at any node $i$ on the power grid, over the distribution of the noise injected at one or several nodes. 
Non-Gaussianities are quantified by nonzero third and fourth cumulants -- skewness and kurtosis. We are particularly interested in finding how far they 
propagate away from the noise sources, and what is their fate in the presence of multiple independent sources of noise.
First, we find that non-Gaussianities
in noise disturbances propagate differently, depending on 
the relation between $\tau_0$ and the intrinsic network timescales. When $\tau_0$ is the shortest
time scale, non-Gaussianities disappear over short distances relative to Gaussian fluctuations, while when $\tau_0$ is the longest time scale, 
they propagate through the whole system just like Gaussian fluctuations do, independently of the distribution of inertia. This is what happens when a single noise source is present.
Second, from a Berry-Esseen theorem we show that, 
for identically but independently distributed sources of noise, non-Gaussianities disappear with the number of noise sources in both 
asymptotics of short and long $\tau_0$. Our analytical results are corroborated by numerical simulations on realistic power grids. 
Compared to earlier works on noise propagation in complex synchronous
networks of oscillators and power grids~\cite{Wil06,Wit15,Sia16,Ket16,Yan17,Hae18,Pag19,Tyl19,Wol19,Sch20,Dev12,Tyl18c,Hin19},
our manuscript (i) goes beyond the white-noise limit, and 
includes in particular regimes of long noise correlation time that are particularly relevant for high-voltage power networks, (ii) is based on analytical calculations, and (iii) considers
the case of multiple sources of
power feed-in noise. Our approach relies on a single restrictive assumption, that the non-Gaussianities can be modelled by the first few 
cumulants of their distribution. While this excludes Lorentzian and power-law distributions with small exponents, it is not an
important restriction, however, since the frequency fluctuations that have been reported in power systems so far
exhibit close to exponential tails~\cite{Hae18,Ryd20,Ryd21,Sch18}. 

The manuscript is organized as follows. Following this introduction, we construct our dynamical model for the dynamics of
high-voltage AC power grids in Section~\ref{sec2}. 
Analytical results are derived and presented in Section~\ref{sec3}. We confirm them numericallly
in Section~\ref{sec4} and discuss their importance and relevance in Section~\ref{sec5}.

\section{The Dynamical Model}\label{sec2}

\subsection{The Swing Equations} 

The operational state of an AC power grid is determined by complex voltages $V_i = |V_i| \exp[i \phi_i]$ at each of the $i=1, \ldots N$ nodes of the grid. In normal operation, voltage amplitudes are fixed not far from their rated value, and voltage angles rotate close to the rated
frequency, $\phi_i(t)= \Omega_0 t + \theta_i(t)$, with $\Omega_0/2 \pi = 50$ or 60Hz. Over time intervals ranging roughly 
from seconds to several tens of seconds, the transient dynamics of high-voltage power grids is given by 
the swing equations~\cite{Mac08}. They govern the time-evolution of voltage angles $\theta_i(t)$ in a frame rotating at the rated frequency.
In high-voltage power grids, a standard approximation is
the lossless line approximation, which neglects Ohmic losses. The swing equations then read
\begin{equation}\label{eq:generators}
    m_i \ddot{\theta}_i + d_i \dot{\theta}_i = P_i - \sum_{j}B_{ij}\sin(\theta_i-\theta_j) \, ,
\end{equation}
with the inertia ($m_i$) and damping ($d_i$) parameters. The active power $P_i$ is positive for generators and negative for loads, 
and $B_{ij}=b_{ij} |V_i| |V_j|$ denotes the product of the voltage magnitudes at nodes $i$ and $j$ with
the line susceptance. In the lossless line approximation, line conductances are neglected. 

At equilibrium, electric power grids are synchronized network systems~\cite{Are08}. They lie close to an operational synchronous 
state where voltage angles are solutions to 
a set of transcendental equations called the power flow equations. Under our assumptions, they read
\begin{equation}\label{eq:powerflow}
P_i = \sum_{j}B_{ij}\sin(\theta_i-\theta_j) \, .
\end{equation}
Their solution $\{\theta^{(0)}_i\}_{i=1, \ldots N}$ corresponds to the instantaneous, synchronous operational state of the power grid.

\subsection{Wave Propagation}

We want to investigate how a local perturbation about the solution to Eq.~\eqref{eq:powerflow} propagates across the system
and influences voltage angles far away from it.
Such an occurence is illustrated for the PanTaGruEl~\cite{Tyl19a} model of the synchronous grid of continental Europe
in Fig.~\ref{fig:RoCoF_snapshots}. Initially, the system is in a steady-state solution of Eq.~\eqref{eq:powerflow}. An 
abrupt power loss $P_i \rightarrow 0$, corresponding to the disconnection of a large power plant in Spain, brings the system out of equilibrium.
Following that perturbation, a voltage-angle wave propagates across the grid, which is represented in five consecutive 
color-coded snapshots in Fig.~\ref{fig:RoCoF_snapshots}. 

In large power grids, even the loss of large power plants is a relatively weak perturbation in a mathematical sense.
For instance the European Network of Transmission System Operators for Electricity (ENTSO-E) reference incident
considers the simultaneous tripping of two of the largest power plants, connected to the same bus~\cite{reference}. 
This corresponds to less than one percent of the total power injected 
in the synchronous grid of continental Europe. 
For the case plotted in Fig.~\ref{fig:RoCoF_snapshots} 
of a power loss of $\Delta P=900$MW, corresponding to a large power plant, frequency deviations
never exceed $2 \pi \Delta \omega =  0.12$ Hz, i.e. a fraction of a percent of the rated frequency~\cite{Pag19}. 
Power feed-in noises being by nature smaller than the rated power on which they are superimposed, 
it is therefore legitimate to investigate them through the linearization of
Eq.~\eqref{eq:generators} about the operational synchronous state. With
$\theta_i = \theta^{(0)}_i + \delta\theta_i$ and $P_i = P_i^{(0)} + \delta P_i$, one gets
\begin{equation}\label{eq:swinglin}
    \mathbf{M} \delta \ddot{\bm{\theta}} + \mathbf{D} \delta \dot{\bm{\theta}} = \delta \mathbf{P} - \mathbf{L} \delta \bm{\theta} \, ,
\end{equation}
where we grouped the voltage angle deviations into a vector $\delta \bm \theta$, and introduced the diagonal inertia and damping matrices,
$\mathbf{M} = \diag(m_i)$ (with $m_i=0$ on load nodes), $\mathbf{D} = \diag(d_i)$ as well as the weighted network Laplacian matrix $\mathbf{L}$,
\begin{equation}
    \mathbf{L}_{ij} = \begin{cases}
        -B_{ij} \cos(\theta_i^{(0)} - \theta_j^{(0)}) \, ,    & \text{for } i \neq j \, , \\
        \sum_{k}B_{ik}\cos(\theta_i^{(0)}-\theta_j^{(0)}) \, , & \text{for } i = j \, .
    \end{cases}
\end{equation}
The perturbation generating a wave of voltage angle and frequency disturbances is encoded in 
the source term vector $\delta \mathbf{P}$, whose components are non-zero at nodes where the perturbation is active. 
Below we consider cases of (i) a single noisy perturbation and (ii) a collection of independent, geographically distributed noisy perturbations.

Power grids have two types of nodes, corresponding to power plants and loads. They have very different dynamical inertia and damping parameters. 
Most loads as well as inverter-connected, new renewable sources of energy have no inertia, $m_i=0$, while traditional power plants
have an inertia roughly proportional to their rated power output~\cite{Mac08}. Furthermore, 
loads have a damping parameter significantly smaller than generators~\cite{Ber81}. While it is crucial to incorporate these dynamic inhomogeneities in 
any analysis of realistic power grids, they render analytical approaches intractable. Recent works, took a perturbation theory approach to incorporate 
small deviations about the homogeneous case~\cite{Pag19a,Fri22}, however most are based on homogeneity
assumptions~\cite{Teg15,Pag17,Poo17,Gru18}. To justify it, one often invokes a 
Kron-reduction of the network~\cite{Kro39}
into an effective network with modified line susceptances connecting only inertiaful, generator nodes. This transformation 
is based on Schur's complement formula~\cite{Hor86}, and since the reduced load nodes have no inertia and a much smaller damping term, 
this reduction modifies the dynamics on the generators only marginally. Once the reduction is performed, one furthermore argues that 
considering uniform damping and inertia, $d_i=d$, $m_i=m$ is justified, because, only large plants, all with
with large rated power are connected to the high-voltage grids we are focusing on here. Additionally, machine
measurements indicate that the ratio of damping over inertia does not vary by much from one machine to another~\cite{gamma}. 
Hence, in our analytical treatment, we consider noise propagation
from Eq.~\eqref{eq:swinglin} for a Kron-reduced network with homogeneous dynamic parameters, $d_i=d$, $m_i=m$. 
However, our numerical treatment is entirely free from this assumption and is based on a realistic, inhomogeneous power-grid model. 
Our numerical results validate our analytical results in the particularly relevant regime of long noise correlation time. 

\section{Disturbance wave propagation : Analytical approach}\label{sec3}

\subsection{Linearized swing equations and modal decomposition}\label{sec:Swing}

Eq.~\eqref{eq:swinglin} is a damped wave equation with a source term. It is defined on a discrete, meshed complex network encoded
in the Laplacian matrix $\mathbf{L}$, which accordingly replaces the Laplace operator of continuous wave equations. Given a source term, 
we compute the moments $\mu_p \equiv \langle \delta \theta_i^p(t \rightarrow \infty) \rangle$, $p \le4$ of the distribution of angle deviations at any node $i$ on the network. Here, $t\rightarrow \infty$ means that the observation takes place long after the onset of the noisy perturbation, to avoid transient behaviors.
To do so, we use a modal expansion of 
Eq.~(\ref{eq:swinglin}) over the set of eigenmodes $\{ \mathbf{u}_{\alpha} \}$ 
of the Laplacian matrix $\mathbf{L}$. Writing $\delta\theta_i(t)=\sum_\alpha c_\alpha(t) \, u_{\alpha,i}$\,, Eq.~(\ref{eq:swinglin}) becomes
\begin{eqnarray}\label{eq:langevin2}
m\,\ddot{c}_\alpha + d\,\dot{c}_\alpha + \lambda_\alpha c_\alpha =  \delta \mathbf{P}(t) \cdot \mathbf{u}_{\alpha}  \,,
\end{eqnarray}
where $ \mathbf{L} {\bf u}_{\alpha}= \lambda_\alpha {\bf u}_{\alpha}$, with $\lambda_\alpha\ge 0$, $ \alpha=1, \ldots N$. Eq.~\eqref{eq:langevin2} is 
the differential equation for a damped, driven harmonic oscillator. It is 
easily solved by means of Laplace transforms. The general solution reads~\cite{Tyl19a}
\begin{eqnarray}\label{eq:calpha}
c_{\alpha}(t)&=& m^{-1}e^{-(\gamma+\Gamma_{\alpha})t/2}\int_0^{t}e^{{\Gamma_{\alpha}}t_2}  \nonumber \\
&&\times \int_{0}^{t_2} \, e^{(\gamma-\Gamma_{\alpha}) t_1 /2} \, \delta \mathbf{P}(t_1) \cdot \mathbf{u}_{\alpha} \,  \, {\rm d}t_1{\rm d}t_2 \;,\end{eqnarray}
with $\Gamma_\alpha=\sqrt{\gamma^2-4\lambda_\alpha/m}$ and $\gamma=d/m$\,. 
Moments $\mu_p$ 
of voltage-angle deviations are calculated as averages over the noise distribution. From Eq.~\eqref{eq:calpha}, $\mu_p$  
contains an average $\langle  \delta P_{i_1} (t_1)\delta P_{i_2} (t_2) \ldots \delta P_{i_p} (t_p)\rangle$, 
over the product of $p$ sources of noise inside exponential integrals. One therefore needs to specify the moments
of the noise distribution. We start from a geographically uncorrelated feed-in noise on nodes labeled $i_0$, 
whose first two moments are given by
\begin{subequations}\label{eq:moments}
\begin{eqnarray}
\langle \delta P_{i_0}(t_1) \rangle &=&0 \, ,\\
\label{eq:2pcorr}
\langle \delta P_{i_0}(t_1) \delta P_{i_0}(t_2) \rangle &=& \sigma^2  \,
e^{- |t_1-t_2|/\tau_0} \, , 
\end{eqnarray}
\end{subequations}
to which we add non-Gaussianities in the skewness and kurtosis of the noise distribution as
\begin{widetext}
\begin{subequations}\label{eq:moments}
\begin{eqnarray}
\label{eq:3pcorr}
\langle \delta P_{i_0}(t_1) \delta P_{i_0}(t_2) \delta P_{i_0}(t_3) \rangle &=& a_3 \, \sigma^3   \, \prod_{m<n} 
e^{-|t_{i_m}-t_{i_n}|/\tau_0}  \, , \\
\label{eq:4pcorr}
\langle \delta P_{i_0}(t_1) \delta P_{i_0}(t_2) \delta P_{i_0}(t_3) \delta P_{i_0}(t_4) \rangle_c &=& a_4 \,
\sigma^4 \,  \, \prod_{m<n} e^{-|t_{i_m}-t_{i_n}|/\tau_0}  \, ,
\end{eqnarray}
\end{subequations}
\end{widetext}
where $\langle \ldots \rangle_c$ explicitely refers to the cumulant. This in particular substracts all
disconnected averages such as  $\langle \delta P_{i_0}(t_1) \delta P_{i_0}(t_2)\rangle\langle \delta P_{i_0}(t_3) \delta P_{i_0}(t_4) \rangle$. 
The parameters $a_{3,4}$ characterize non-Gaussianities in the noise distribution. They correspond to skewed distributions
($a_3 \ne 0$), with tails longer ($a_4 >0$) or shorter ($a_4 <0$) than the normal distribution.

The moments $\mu_p$ are given by exponential integrals and are straightforwardly calculated. However, their exact expressions are somewhat
complicated. We give them for the variance only and, for the third and fourth cumulants, 
discuss limiting cases of long and short correlation time. 

\subsection{Time scales in high-voltage power grids vs. noise correlation time}

Eq.~\eqref{eq:langevin2} makes it clear that,
beside $\tau_0$, the other time scales are the damping time $\gamma^{-1}=m/d$, the $\alpha^{\rm th}$ oscillator period 
$T_\alpha=\sqrt{m/\lambda_\alpha}$ and the combination $\gamma T_\alpha^2= d/\lambda_\alpha$ of the two~\cite{Tyl19}.
For the synchronous grid of continental Europe, a detailed analysis based on realistic line admittances and dynamic parameters in large-scale
power grids gave estimates for these time scales as
$\gamma^{-1} \simeq 2.5 s$, $T_\alpha < 1 s$ and $\gamma T_\alpha^2 < 0.4 s$ $\forall 
\alpha$. Therefore the regime of long noise correlation time is already reached for $\tau_0 \gtrsim 5-10 s$,
while the short correlation time regime requires $\tau_0 \lesssim 1 \mu s$~\cite{Pag19,Tyl19}.

While circuit breakers and other switches may disconnect power lines and put power plant off-line in a fraction of a second, 
disconnection-reconnection sequences may occur at most two to three times consecutively by design of these switches. 
It is hard to think of a significant noise perturbation 
fluctuating persistently on a time scale shorter than a few seconds. Moreover, in the Supplemental Material we show several examples of feed-in power fluctuations from renewables with $\tau_0 \gtrsim 2-5$ mins.
Hence, we conclude that the long correlation time regime is the relevant one for
high-voltage transmission grids we are interested in.  

We first consider the case of a single source of noise and discuss multiple noisy nodes in paragraph \ref{SecE}.

\subsection{Voltage angle variance}

In the limit of large observation time, the voltage angle variance is given in Eq.~(S8) of the Supplemental Material.
The two limiting cases of long and short noise correlation time $\tau_0$ give
key insights on noise propagation. 

First, when $\tau_0$  is the largest time scale, 
the voltage angle variance at node $i$ reads
\begin{eqnarray}\label{eq:varlong}
\lim_{\tau_0 \rightarrow \infty} \langle \delta\theta_i^2 \rangle &=& \left( \sigma \sum_{\alpha}\frac{u_{\alpha,i_0}u_{\alpha,i}}{\lambda_\alpha}\right)^2\, .
\end{eqnarray}
The quantity squared inside the parenthesis is the Green's function for the linear operator $\mathbf{L}$, from the noise source to the observation node $i$. 
For optical or electronic waves propagating through disordered mesoscopic systems, 
quantities similar to $\langle \delta\theta_i^2 \rangle$ in Eq.~\eqref{eq:varlong}
decay as power laws with the distance between $i_0$ and $i$, when averaged 
over a relatively narrow but high-lying spectral interval~\cite{Akk07}. Eq.~\eqref{eq:varlong} instead corresponds to a ``zero-energy" Green's
function, indicating that fluctuations with long correlation times are transmitted by a few low-lying, long-wavelength eigenmodes of $\mathbf{L}$,
for which the perturbative approaches of Ref.~\cite{Akk07} cannot be directly applied. 

Second, when $\tau_0$ is the shortest time scale, one obtains
\begin{eqnarray}\label{eq:varshort}
\lim_{\tau_0 \rightarrow 0}
\langle \delta\theta_i^2 \rangle &=& 2 \sigma^2 \tau_0 \sum_{\alpha,\beta \ge 2}\frac{u_{\alpha,i_0}u_{\beta,i_0}u_{\alpha,i}u_{\beta,i}}{d(\lambda_\alpha + \lambda_\beta) + \frac{m}{2d}(\lambda_\alpha - \lambda_\beta)^2} \, . \qquad
\end{eqnarray}
In the inertialess limit, $m=0$, the variance is given by a two-particle Green's function. 

\subsection{Higher voltage angle cumulants : Long correlation time regime}\label{sec:3c}

In the limit of long correlation time, it is straightforward to show that higher cumulants behave similarly to the variance, Eq.~\eqref{eq:varlong}, namely 
\begin{eqnarray}\label{eq_moments}
\lim_{\tau_0 \rightarrow \infty} \langle \delta \theta_i^p \rangle_c  &=& a_p \left( \sigma \sum_{\alpha}\frac{u_{\alpha,i_0}u_{\alpha,i}}{\lambda_\alpha}\right)^p  , \qquad
\end{eqnarray}
with the parameters $a_p$ giving the deviation from Gaussianity in the feed-in fluctuations, Eqs.~\eqref{eq:moments}.
The most remarkable thing is that, from Eqs.~\eqref{eq:varlong} and \eqref{eq_moments}, 
standardized higher cumulants are given by
$\langle \delta \theta_i^p \rangle_c / \langle \delta \theta_i^2 \rangle^{p/2}=a_p$, regardless of the distance between the measured node and the 
noise source. This is one of the
main results of this paper: long correlation times boost non-Gaussian 
fluctuations from a single noise source so much, that they propagate and persist 
over the whole network. This result is in particular independent of inertia, suggesting, as previously found in Ref.~\cite{Tyl21css}, that 
disturbances with long characteristic times are affected only marginally by inertia, even when the latter is inhomogeneous. Below, we 
confirm the validity of this conjecture with numerical simulations on realisticially inhomogeneous power grids corroborating our theoretical predictions.

\subsection{Higher voltage angle cumulants : Short correlation time regime}

The result for the third moment is given in Eq.~(S9) in the Supplemental Material.
In the limit of vanishing inertia, $m=0$, it gives,
\begin{eqnarray}\label{eq:kurts}
\lim_{\tau_0 \rightarrow 0}
\langle \delta\theta_i^3 \rangle &=& \sigma^3 \tau_0^2 \sum_{\alpha,\beta,\gamma\ge 2}\frac{u_{\alpha,i_0}u_{\beta,i_0}u_{\gamma,i_0}u_{\alpha,j}u_{\beta,j}u_{\gamma,j}}{d^2 (\lambda_\alpha+\lambda_\beta+\lambda_\gamma )} \, , \qquad
\end{eqnarray}
which, together with Eq.~\eqref{eq:varshort}, reflects the fact that, for Kuramoto, i.e. inertialess oscillators, non-Gaussianities in 
the $p^{\rm th}$ cumulants propagate as a $p-$particle Green's function in the white-noise limit.

\begin{figure}
\includegraphics[width=0.45\textwidth]{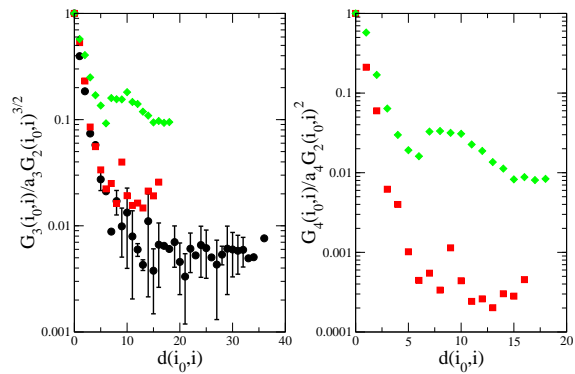}\vspace{-1.5cm}
\caption{Left panel: Normalized ratio $G_3/a_3 G_2^{3/2}$ of the $3-$particle Green's function and the $3/2-$power of the two-particle Green's function.
Right panel: Normalized ratio $G_4/a_4 G_2^{2}$ of the $4-$particle Green's function and the square of the two-particle Green's function. In the 
inertialess case, these ratios give the standardized skewness and kurtosis of the voltage angle fluctuations in the case of short noise correlation time.
 Black symbols correspond to the full PanTaGruEl model of the synchronous grid of continental Europe~\cite{Pag19,Tyl19},
 red ones to a connected subsection with N=1000 nodes of PanTaGruEl and green ones to the SciGRID model of the 
 high voltage power grid of Germany~\cite{scigrid}. PanTaGruEl data for $G_4$ are missing because they require prohibitively large computation times. 
}\label{figG}
\end{figure}

As mentioned previously, Green's functions decay exponentially with distance in disordered mesoscopic systems, however it is not clear whether this 
behavior applies to the "zero-energy" case considered here, nor to $p$-particle Green's functions. To understand better the propagation of non-Gaussianities in 
the short correlation time regime, we therefore numerically evaluate the expressions in 
Eqs.~\eqref{eq:varshort} and \eqref{eq:kurts}. Fig.~\ref{figG} shows the theoretically predicted standardized skewness and kurtosis 
(normalized by $a_3$ and $a_4$ respectively) according to Eqs.~\eqref{eq:varshort} and \eqref{eq:kurts}. We consider
inertialess networks, and plot the results as a function of the
geodesic distance to the source of noise, for various power-grid models. In contrast to the 
the long correlation time prediction, both skewness and kurtosis decay fast away from the noisy node (seemingly exponentially fast) before they saturate at a 
constant, small value. 
We conclude that short noise correlation times suppress the propagation of non-Gaussianities
through meshed networks over a network-dependent distance.

\subsection{Multiple sources of noise}\label{SecE}

We finally consider the case with $M$ distinct, independently but identically distributed sources of power feed-in fluctuations. 
In that case, there are $M$ contributions similar to that in Eq.~\eqref{eq_moments} to the cumulant, but $M! (M-p/2)!/(p/2)!$
pairings of the noise sources for the moment of even order $p$. These latter contributions are much more numerous and they 
result in a Gaussian $p^{\rm th}$ moment -- this is the standard mechanism behind central limit and Berry-Esseen theorems~\cite{Durrett}. 
For instance for the $p=4$ moment in the large correlation time limit, one obtains
\begin{eqnarray}\label{eq_momentsM}
\lim_{\tau_0 \rightarrow 0} &&
\langle \delta \theta_i^4 \rangle =\sum_{i_0=1}^M \Big(\sigma \sum_{\alpha}\frac{u_{\alpha,i_0}u_{\alpha,i}}{\lambda_\alpha}\Big)^4  \nonumber \\
&&+ 3\sum_{i_0 < j_0} \Big(\sigma \sum_{\alpha}\frac{u_{\alpha,i_0}u_{\alpha,i}}{\lambda_\alpha}\Big)^2 \Big(\sigma \sum_{\beta}\frac{u_{\beta,j_0}u_{\beta,i}}{\lambda_\beta}\Big)^2 , \qquad
\end{eqnarray}
where the factor $3$ in the second line accounts for all possible pairings between the product of four noise sources, $\delta P_{i_l}$, $l=1, \ldots 4$. The pairing mechanism giving
the second term on the
right-hand side of Eq.~\eqref{eq_momentsM} leads to 
the convergence of the voltage angle distribution to a Gaussian distribution, with $\langle \delta \theta_i^4 \rangle/\langle \delta \theta_i^2 \rangle^2 \rightarrow 3$ [to see this, sum over the noisy nodes $i_0=1, \ldots M$ in Eq.~\eqref{eq_moments} and compare the result with 
\eqref{eq_momentsM}]. The convergence is the same as in the Berry-Esseen theorem~\cite{Durrett}.
When $M$ is large, this second term  dominates over the first one by a factor $(M-1)/2$, so that the ratio of 
the fourth cumulant -- a measure of non-Gaussianity --  to the fourth moment becomes $\propto M^{-1}$. With the standard 
definition~\cite{Durrett}, non-Gaussianities disappear at a rate $\propto M^{-1/2}$.

\section{Numerical Results}\label{sec4}

Our two main theoretical predictions are that, 

(i) non-Gaussianities propagate over the entire network when they originate from a noisy source with long
correlation time, while they disappear exponentially with the distance from the source for short noise correlation time, 

(ii) non-Gaussianities 
become smaller with the number $M$ of uncorrelated sources of noise. We confirm these two predictions by numerical integration of Eq.~\eqref{eq:generators}
for various networks with single or multiple sources of noise with short and long correlation times. 

\begin{figure}
\includegraphics[width=0.47\textwidth]{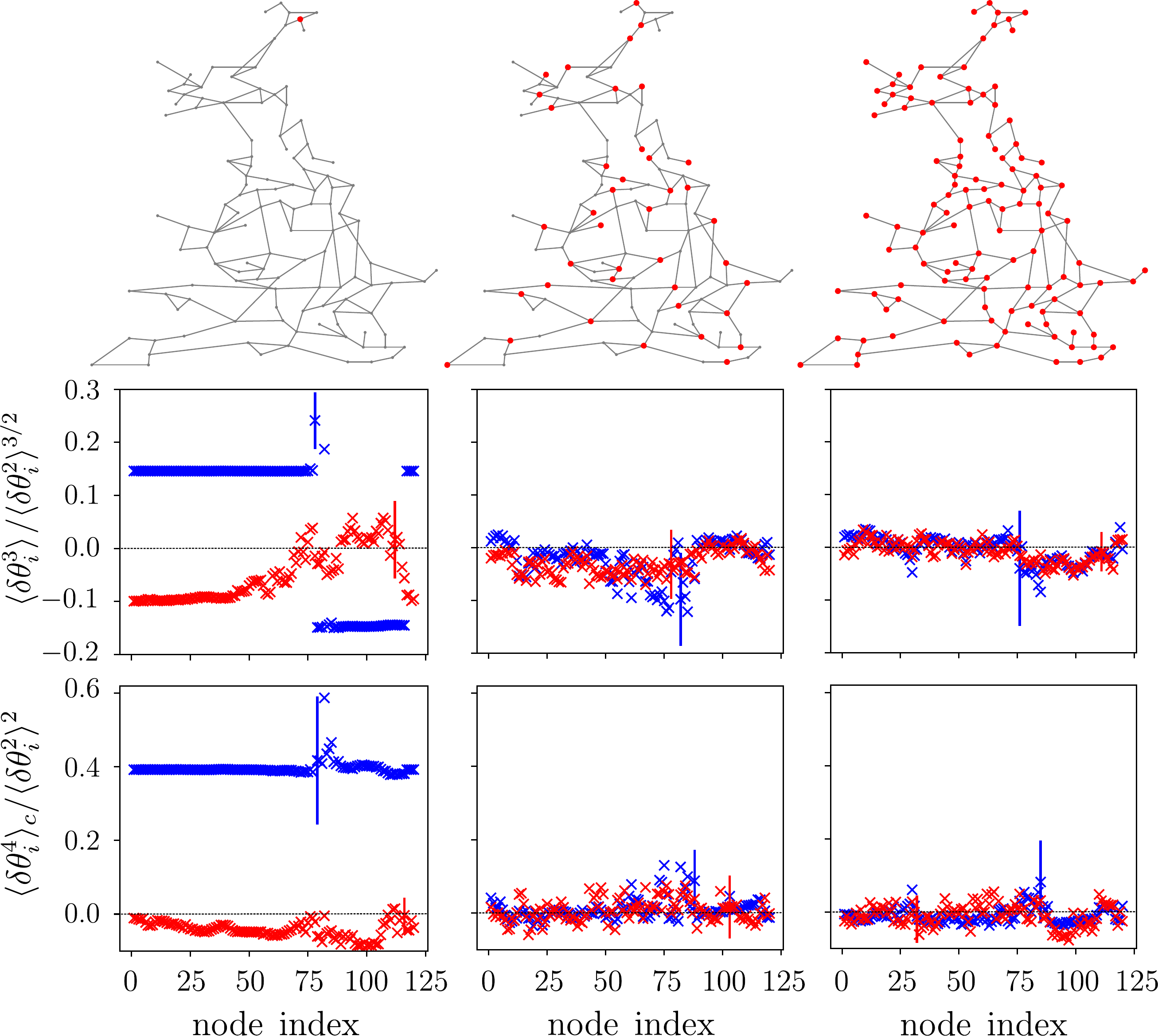}
\caption{Numerically evaluated voltage-angle skewness and kurtosis for the UK grid with 1 (left column), 40 (middle) and 120 (right)
sources of noise, whose locations are shown in the grid map in the top panels. Blue (red) crosses correspond to noise with long (short) correlation time. 
Error bars are indicated on nodes that have the largest numerical fluctuation when doubling the simulation time. Injected noises have skewness and kurtosis with
$a_3=-0.15$ and $a_4=0.4$ for long correlation time and $a_3=-1.4$ and $a_4=18.9$ 
for short correlation time.}\label{fig3}
\end{figure}

Fig.~\ref{fig3} first illustrates how voltage-angle fluctuations behave as more uncorrelated sources of noise are added. Blue and red crosses correspond to long and short
noise correlation times respectively, and three situations of a single (left column), 40 (middle) and 120 (right) uncorrelated sources of noise are shown.
When a single noise source is present, skewness and kurtosis of voltage-angles directly reflect their value for the noise source for long noise correlation
time, while they are significantly reduced for short correlation time. In the latter case, a finite skewness persists over more than half of the network, and fluctuates
about zero for the rest of the network nodes. As the number of noise sources increases, both skewness and kurtosis are suppressed as
voltage-angles become normally distributed following the action of the Berry-Esseen theorem. Numerical data still fluctuate due to the discreteness of 
time steps and the finiteness of the integration time. Error bars indicate the largest data variation upon doubling of the integration time. 
A remarkable feature is the sign change in the long-correlation-time skewness in the middle-left panel.
It is easily understood when re-expressing the single-particle Green's function in terms of graph theoretic indicators as
\begin{align}\label{eq:grth}
\begin{split}
\sum_{\alpha}\frac{u_{\alpha,i_0}u_{\alpha,i}}{\lambda_\alpha}=-\frac{1}{2}[\Omega_{i_0i}-C_1^{-1}(i_0)-C_1^{-1}(i) +2\Kf_1/n^2]\,,
\end{split}
\end{align}
where $\Omega_{i_0,i}$ is the resistance distance between node $i_0$ and $i$~\cite{Kle93}, $C_1(i)=(n^{-1}\sum_j\Omega_{ij})^{-1}$ is the resistance centrality of node $i$ and $\Kf_1=\sum_{i<j}\Omega_{i,j}$ is the so-called Kirchhoff index. The sign of all odd-$p$ cumulants in the long correlation time limit, see Eq.~\eqref{eq_moments}, is given by the $p^{\rm th}$ power of the Green's function. It is therefore determined by a trade-off 
between the centralities of the input and measured nodes on the one hand, and the resistance distance between them on the other hand.
As but one consequence, the skewness changes sign as the  measurement point $i$ 
is taken further and further away from $i_0$, when the resistance distance $\Omega_{i_0i}$ becomes larger than the sum of the inverse 
node centralities in Eq.~\eqref{eq:grth}. Theory-simulation agreement for the UK model 
with homogeneous damping and inertia parameters is excellent, and well within the error bars of finite-time integration.

\begin{figure}
\includegraphics[scale=0.55]{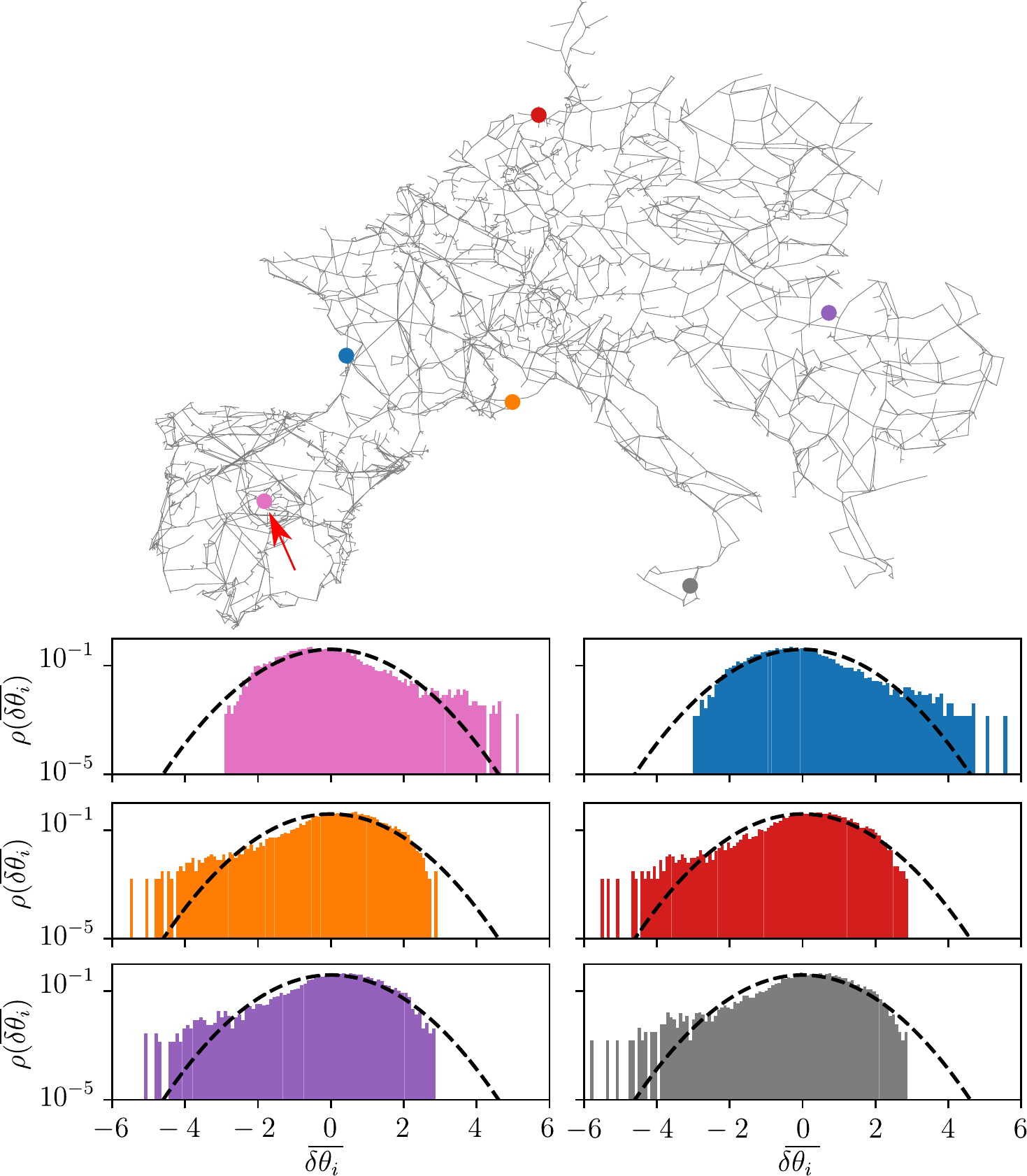}
\caption{Normalized distributions of voltage-angles $\overline{\delta\theta_i}=\delta\theta_i/\sqrt{\langle \delta\theta_i^2 \rangle}$, from a single non-Gaussian source of noise for the PanTaGruEl model of the
synchronous grid of continental Europe with constant inertia and damping~\cite{Pag19,Tyl19}. 
The source of noise is located at the pink node and indicated by the red arrow in the top panel.
The noise is in the regime of long correlation time. 
Voltage angle deviations are measured at the five other colored nodes. All normalized distributions are the same, up to a
sign inversion, in agreement with our analytical prediction of Eq.~\eqref{eq_moments}.
Dashed lines indicate a Gaussian distribution.}\label{fig4}
\end{figure}

We next turn our attention to a larger-scale, more realistic model of a high-voltage power grid and consider the PanTaGruEl model of the 
synchronous grid of continental Europe~\cite{Pag19,Tyl19}. As discussed in Section~\ref{sec:Swing}, intrinsic time scales in such large-scale power grids  
are such that  the short correlation time limit corresponds to $\tau_0 \lesssim 1 \mu s$ while the long correlation time regime corresponds to 
$\tau_0 \gtrsim 5-10 s$. Persistent sources of noise therefore correspond to the long correlation time regime, which we focus on.

\begin{figure}
\includegraphics[scale=0.55]{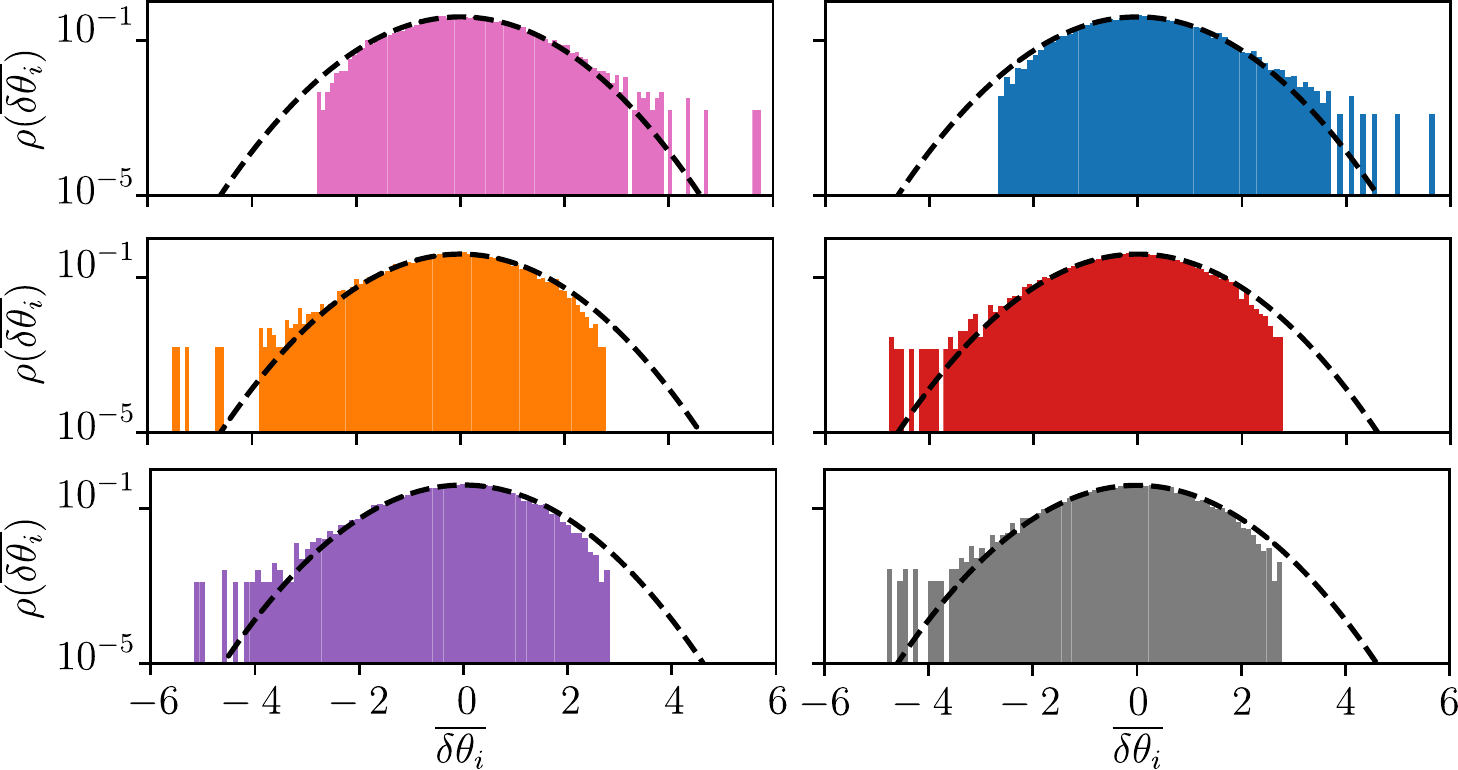}
\caption{Normalized distributions of voltage-angles $\overline{\delta\theta_i}=\delta\theta_i/\sqrt{\langle \delta\theta_i^2 \rangle}$, as in Fig.~\ref{fig4}, but for an inhomogeneous, realistic configuration of inertia and damping in the PanTaGruEl
model of the synchronous grid of continental Europe~\cite{Pag19,Tyl19}.
}\label{fig5}
\end{figure}

Fig.~\ref{fig4} shows data for  a 
homogeneous power grid with constant inertia and damping, $m_i=m$, $d_i=d$ in Eq.~\eqref{eq:generators}, $\forall i$. 
A non-Gaussian power-feed in noise is injected at the pink node indicated by the red arrow on the network map, and
voltage-angle fluctuations are measured at the five other colored nodes. One sees first, that all voltage-angle distributions are the same, up to 
a sign inversion $\delta \theta_i \rightarrow -\delta \theta_i$. This corroborates our prediction of Eq.~\eqref{eq_moments}, according to which 
all standardized cumulants are the same, up to possible sign changes in odd cumulants, in the case of noise with long correlation time. The observed sign change is 
consistent with Eq.~\eqref{eq:grth}, where the blue node has the same normalized voltage-angle distribution as the source, pink node, because it is close to it
and the right-hand side in Eq.~\eqref{eq:grth} is dominated by the sum of the inverse centralities, $C_1^{-1}(i_0)+C_1^{-1}(i)>\Omega_{i_0,i}$.
All other nodes are further away and correspond to a regime where the inequality is reversed,  $C_1^{-1}(i_0)+C_1^{-1}(i)<\Omega_{i_0,i}$ and 
odd cumulants undergo a sign change. 

In the regime of long correlation time, we saw in Section~\ref{sec:3c} that voltage-angle cumulants depend neither on inertia, nor on damping, and following Ref.~\cite{Tyl21css} we conjectured that the prediction of Eq.~\eqref{eq:grth} also applies to cases with inhomogeneous inertia and damping. 
We confirm this conjecture in Fig.~\ref{fig5}, where the normalized voltage-angle distributions 
also keep their non-Gaussianities all over the network, 
{\it regardless of the distance between source and measurement nodes}, in the PanTaGruEl model with realistically inhomogeneous dynamic parameters $m_i$
and $d_i$.

\begin{figure}[h!]
\includegraphics[scale=0.55]{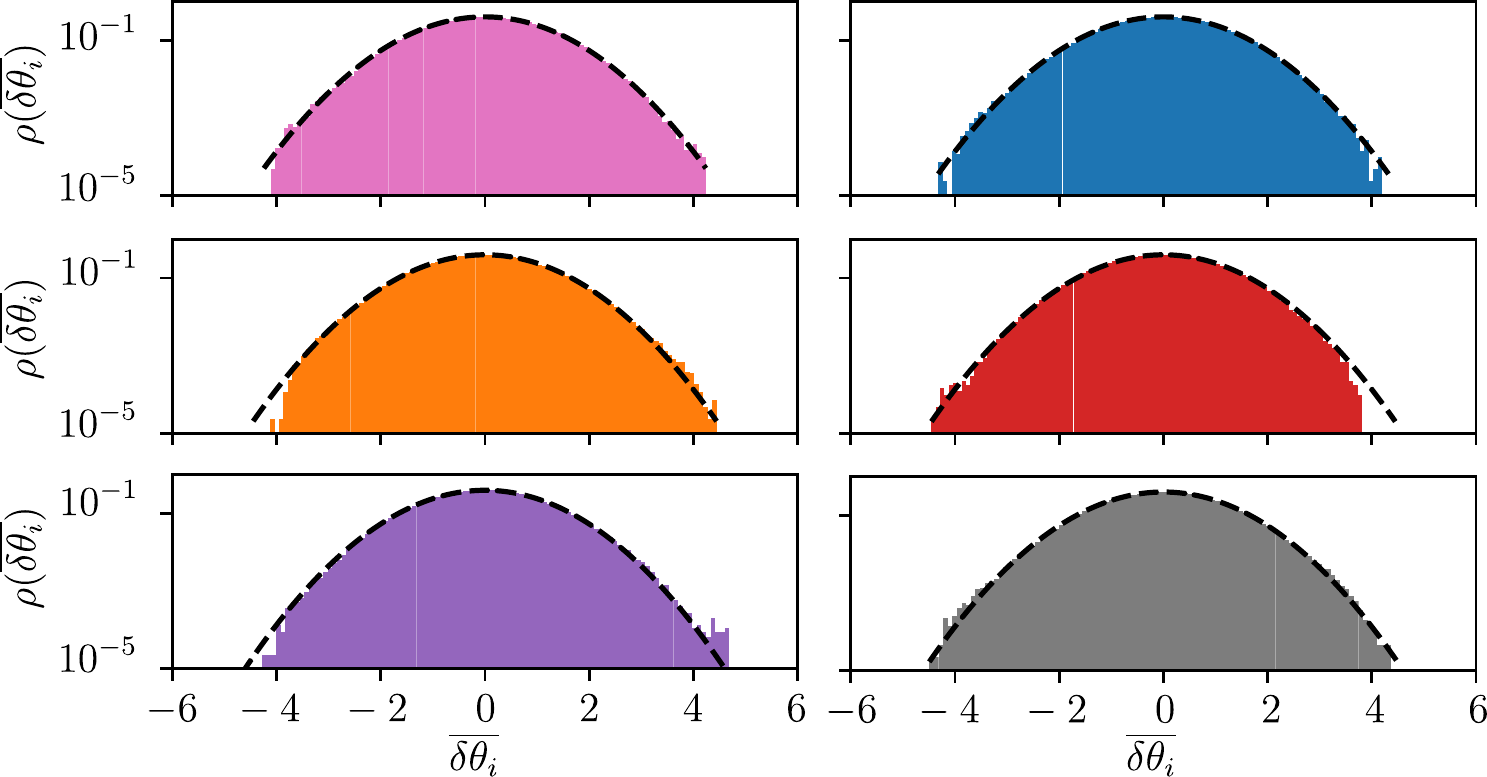}
\caption{Normalized distributions of voltage-angles $\overline{\delta\theta_i}=\delta\theta_i/\sqrt{\langle \delta\theta_i^2 \rangle}$, as in Fig.~\ref{fig5}, but for 381 different, uncorrelated sources of non-Gaussian noise in the PanTaGruEl
model of the synchronous grid of continental Europe~\cite{Pag19,Tyl19}.}\label{fig6}
\end{figure}

Finally, we investigate the case when multiple uncorrelated sources of noise are present. Fig.~\ref{fig6} confirms that, for 381 sources of 
power feed-in fluctuations, non-Gaussianities disappear and voltage-angle deviations become Gaussian distributed. 
In summary, our numerical simulations with realistic power-grid models fully confirm the theoretical predictions presented in Section~\ref{sec3}. \\

\section{Conclusion}\label{sec5}

The theory we just presented has uncovered two previously neglected, yet crucial characteristics that determine how 
voltage-angle disturbances propagate through power grids: 

(i) the correlation time $\tau_0$, i.e. the characteristic time over which sources fluctuate, and 

(ii) the number of sources of fluctuations.

In the white-noise limit of short $\tau_0$, non-Gaussianities decay with the distance from their source and saturate at small values. These values are 
determined by the relevant multi-particle Green's function in the limit of small inertia. 
In the other limit of long correlation times, 
non-Gaussian noise propagates through the whole network, regardless of the distance to the source and independently of inertia, leading to 
voltage angle fluctuations with the same non-Gaussian distribution as the feed-in power noise. 
These non-Gaussianities are, however, averaged out in the presence of multiple, uncorrelated sources of noise. 
 
Modern power grids are rather resilient and in particular able, in a normal operational mode to absorb not too strong voltage angle fluctuations. 
Future grids will be subjected to more disturbances, in particular from new renewable energy sources. A major planification and operational 
concern is that electro-mechanical inertia may be significantly reduced in the future, in particular at times of large renewable power production. 
Our results indicate that, from the point of view of disturbance propagation, one should not be concerned by this reduction of inertia. 

We finally point out that out theory applies generically to diffusively coupled agents, well beyond the power-grid models and similar second-order
Kuramoto models considered in this manuscript. 

\section*{Acknowledgments}

PJ and MT were supported by the Swiss National Science Foundation under grant No. 200020$_-$182050. JH was supported through the U.S Naval Research Laboratory Karles Fellowship.
We thank Julian Fritzsch for discussions on the manuscript, Laurent Pagnier for Fig.~\ref{fig:RoCoF_snapshots} and general discussions, and Will Holmgren and Tucson Electric Power for data on photovoltaic and wind turbine productions.

\pagebreak
\widetext
\begin{center}
\textbf{Finite-time Correlations Boost Non-Gaussian Noise Propagation in High-Voltage Electric Power Grids: Supplemental Information}
\end{center}

\setcounter{equation}{0}
\setcounter{figure}{0}
\setcounter{table}{0}
\makeatletter
\renewcommand{\theequation}{S\arabic{equation}}
\renewcommand{\thefigure}{S\arabic{figure}}
\renewcommand{\bibnumfmt}[1]{[S#1]}
\renewcommand{\citenumfont}[1]{S#1}

\section{Propagation of non-Gaussianities in the Kuramoto Model}

The Kuramoto model corresponds to Eq.~(1) in the main text, with $m_i=0$ and $d_i=1$, $\forall i$.
All cumulants $\langle \delta \theta_i^p \rangle_c$ of angle deviations can be calculated using the
method of Ref.~\cite{melvyn}. Angle deviations are calculated from the linearized differential equations
given in Eq.~(3) of the main text. Expanding
$\delta \theta_i(t) = \sum c_\alpha(t) u_{\alpha i}$ over the eigenvectors ${\bf u}_\alpha=(u_{\alpha1}, \ldots u_{\alpha N})$ of the network Laplacian
$\mathbf L$, one has, with
the initial condition $c_\alpha(t=0) = 0$,
\begin{equation}\label{eq:ca}
c_\alpha(t) = \exp[- \lambda_\alpha t] \, \int {\rm d} t' \exp[\lambda_\alpha t'] \,
\delta {\bf P}(t') \cdot {\bf u}_\alpha \, .
\end{equation}
Cumulants are then straightforwardly calculated from the noise cumulants given in
Eqs.~(7) and (8) in the main text.

\subsection{Angle variance}

We first calculate $\langle \delta \theta^2_i(t) \rangle$. It is straightforward to obtain
\begin{eqnarray}
\langle \delta \theta_i^2(t) \rangle & = & \sum_{\alpha,\beta} \langle
c_\alpha(t) c_\beta(t) \rangle u_{\alpha \, i} u_{\beta \, i} \, ,
\end{eqnarray}
in terms of the components $u_{\alpha, i}$ of the eigenvectors of $\mathbf L$.
Neglecting   terms decaying exponentially with time, one gets
\begin{eqnarray}\label{eq:variance}
\langle \delta \theta_i^2(t) \rangle
& = & \sigma^2 \sum_{\alpha,\beta}
\frac{u_{\alpha \, i_0} u_{\beta \, i_0} u_{\alpha \, i} u_{\beta \, i}}{(\lambda_\alpha+\tau_0^{-1}) (\lambda_\beta+\tau_0^{-1})} + 2 \sigma^2 \tau_0^{-1} \sum_{\alpha,\beta}
\frac{u_{\alpha \, i_0} u_{\beta \, i_0} u_{\alpha \, i} u_{\beta \, i}}{(\lambda_\alpha+\lambda_\beta) (\lambda_\alpha+\tau_0^{-1}) (\lambda_\beta+\tau_0^{-1})} \, .
\end{eqnarray}
The first term in this
expression makes it clear that, at the level of the variance of the angle deviation,
the perturbation propagates as the square of the Green's function $$G_1(i_0,i) = \sum_{\alpha}
\frac{u_{\alpha \, i_0} u_{\alpha \, i}}{\lambda_\alpha} \, , $$
of the Laplacian,  in the limit of long noise correlation times, $\tau_0 \rightarrow \infty$.
In the other limit of short noise correlation time, $\tau_0 \rightarrow 0$,
the perturbation propagates as the two-particle Green's function $$G_2(i_0,i) = \sum_{\alpha,\beta}
\frac{u_{\alpha \, i_0} u_{\beta \, i_0} u_{\alpha \, i} u_{\beta \, i}}{\lambda_\alpha+\lambda_\beta} \, .$$

\subsection{Angle skewness}

A similar calculation gives the third cumulant as
\begin{eqnarray}
\langle \delta \theta^3_i(t) \rangle &=& a_3 \sigma^3 \sum_{\alpha,\beta,\gamma} \,
\frac{u_{\alpha \, i} u_{\alpha \, i_0} u_{\beta \, i} u_{\beta \, i_0}
u_{\gamma \, i} u_{\gamma \, i_0} }{\lambda_\alpha+\lambda_\beta+\lambda_\gamma}
\times \left[ \left(\frac{1}{\lambda_\alpha +2 \tau_0^{-1}}+ \frac{1}{\lambda_\beta + 2 \tau_0^{-1} } \right)
\frac{1}{\lambda_\alpha + \lambda_\beta + 2 \tau_0^{-1} } +
\right. \nonumber \\
&& \left. \;\;\;\;\;\;
\left(\frac{1}{\lambda_\alpha +2 \tau_0^{-1}}+ \frac{1}{\lambda_\gamma + 2 \tau_0^{-1} } \right)
\frac{1}{\lambda_\alpha + \lambda_\gamma + 2 \tau_0^{-1} }
+ \left(\frac{1}{\lambda_\beta +2 \tau_0^{-1}}+ \frac{1}{\lambda_\gamma + 2 \tau_0^{-1} } \right)
\frac{1}{\lambda_\beta + \lambda_\gamma + 2 \tau_0^{-1} }
\right] \, . \qquad \label{eq:skew}
\end{eqnarray}
The two asymptotic limits of large and small correlation time are interesting. First, for $\tau_0 \rightarrow \infty$, one gets
\begin{eqnarray}
\langle \delta \theta^3_i(t) \rangle &=& a_3 \sigma^3 \sum_{\alpha,\beta,\gamma} \, \frac{u_{\alpha \, i} u_{\alpha \, i_0} u_{\beta \, i} u_{\beta \, i_0}
u_{\gamma \, i} u_{\gamma \, i_0} }{\lambda_\alpha   \lambda_\beta \lambda_\gamma} =  a_3 \sigma^3 \left( \sum_{\alpha} \, \frac{u_{\alpha \, i} u_{\alpha \, i_0}}{\lambda_\alpha} \right)^3 = a_3 \sigma^3 G_1^3(i_0,i)\, ,
\end{eqnarray}
i.e. the Green's function $G_1$ to the third power, up to a prefactor.
Taking this result together with that for the second moment, Eq.~\eqref{eq:variance},
it therefore seems to be a general property that cumulants are dominantly given by
the appropriate power of Green's functions in the limit of long noise correlation times. We will confirm this
conjecture below by calculating the kurtosis.

In the other limit of short noise correlation times, $\tau_0 \rightarrow 0$, one obtains
\begin{eqnarray}\label{eq:skew_coupled}
\langle \delta \theta^3_i(t) \rangle &=& (3/2)  a_3 \sigma^3 \tau_0^{2}
\sum_{\alpha,\beta,\gamma} \, \frac{u_{\alpha \, i} u_{\alpha \, i_0} u_{\beta \, i} u_{\beta \, i_0}
u_{\gamma \, i} u_{\gamma \, i_0} }{\lambda_\alpha+\lambda_\beta+\lambda_\gamma} \, ,
\end{eqnarray}
i.e. a three-particle Green's function.
\subsection{Angle kurtosis}

Finally the fourth cumulant of the angle deviations is given by
\begin{eqnarray}
\langle \delta \theta^4_i(t) \rangle &=& a_4 \, \sigma^4 \sum_{\alpha,\beta,\gamma,\delta} \,
\frac{u_{\alpha \, i} u_{\alpha \, i_0} u_{\beta \, i} u_{\beta \, i_0}
u_{\gamma \, i} u_{\gamma \, i_0} u_{\delta \, i} u_{\delta \, i_0} }{\lambda_\alpha+\lambda_\beta+\lambda_\gamma+\lambda_\delta} \nonumber   \\
&& \times \left[
\frac{1}{\lambda_\alpha+3 \tau_0^{-1}}
\frac{1}{\lambda_\alpha+\lambda_\beta+4 \tau_0^{-1}}
\frac{1}{\lambda_\alpha+\lambda_\beta+\lambda_\gamma+3 \tau_0^{-1}}
+ 23 {\rm \; terms \; with \;  index \; permutations}
\right] \, . \nonumber
\end{eqnarray}
The structure of the expression is very similar to Eq.~\eqref{eq:skew}.
In particular it is easily checked that, here again, the $\tau_0 \rightarrow \infty$ limit is dominated by the fourth power of the Green's function $G_1(i_0,i)$, while the other limit $\tau_0 \rightarrow 0$ gives
\begin{eqnarray}\label{eq:kurtosis_coupled}
\lim_{\tau_0 \rightarrow 0} \langle \delta \theta^4_i(t) \rangle &=& (2/3) a_4 \sigma^4 \tau_0^3 \,  \sum_{\alpha,\beta,\gamma,\delta} \,
\frac{u_{\alpha \, i} u_{\alpha \, i_0} u_{\beta \, i} u_{\beta \, i_0}
u_{\gamma \, i} u_{\gamma \, i_0} u_{\delta \, i} u_{\delta \, i_0} }{\lambda_\alpha+\lambda_\beta+\lambda_\gamma+\lambda_\delta} \, ,
\end{eqnarray}
which up to a prefactor is a 4-particle Green's function.
This confirms the above conjecture that $p^{\rm th}$ cumulants of angle fluctuations are given by $p-$particle Green's
functions when the noise correlation time is vanishingly small.

\section{Propagation of non-Gaussianities with inertia}

\begin{figure}
\includegraphics[width=0.6\textwidth]{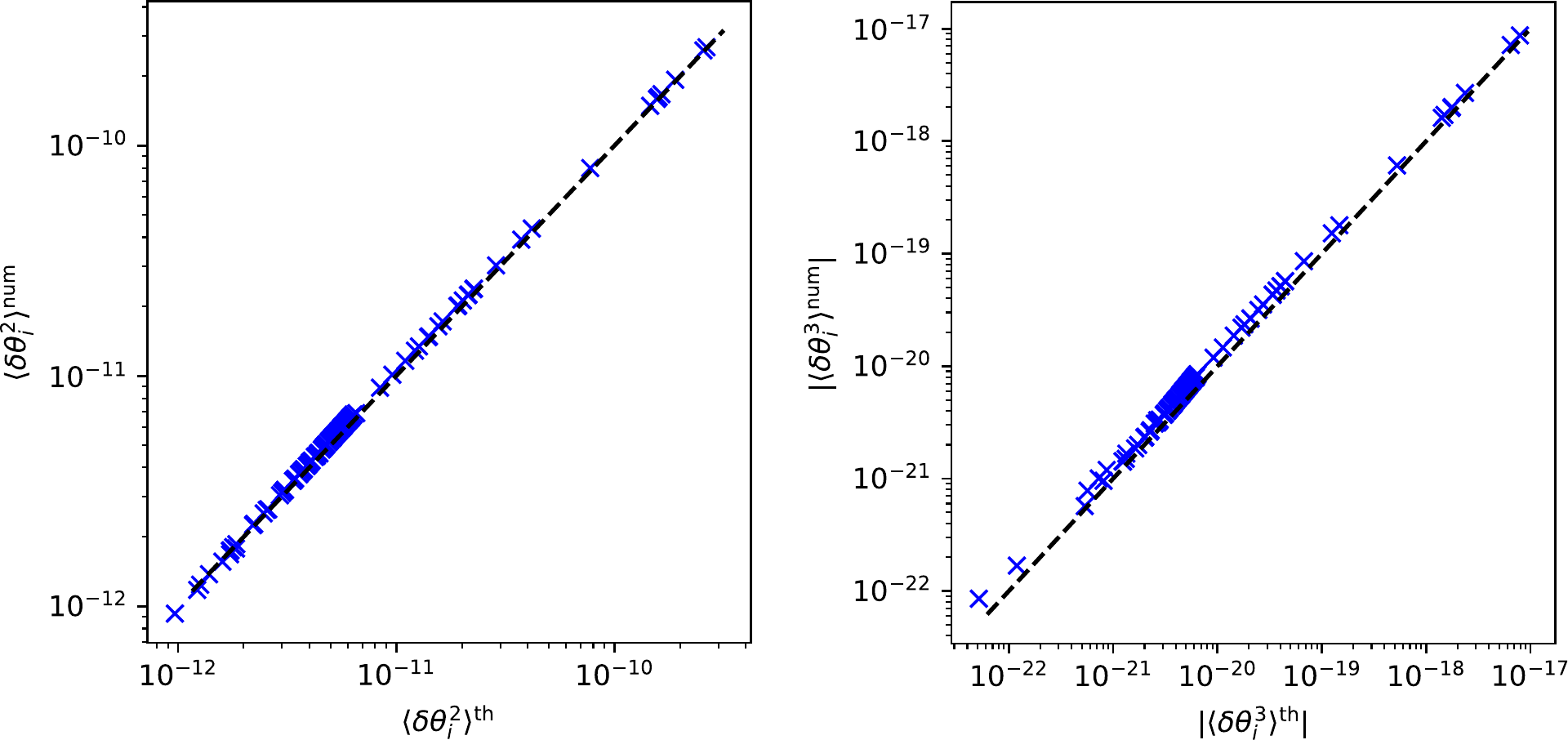}
\caption{Numerically evaluated variance $\langle \delta\theta_i^2 \rangle$ (left panel) and skewness $\langle \delta\theta_i^3 \rangle$ (right panel)
of the angle deviation for each node $i=1, \ldots N$ vs. the theoretical formula,
Eqs.~(\ref{eq_moments_SM}) and \eqref{eq:skewness}, for a single noisy node in the UK grid shown in the top row of Fig.~3 in the main text. The inverse inverse noise correlation time is inside the
spectrum of the network Laplacian, $\lambda_2 < \tau_0^{-1} < \lambda_N$ (left panel) and above the spectrum, $\lambda_N<\tau_0^{-1}$\,.}\label{figS1}
\end{figure}

The same approach allows one to calculate cumulants of angle deviations in the presence of homogeneous inertia terms, with Eq.~(6) in the main
text instead of Eq.~\eqref{eq:ca}.
Instead of Eq.~\eqref{eq:variance}, one obtains for the variance of angle deviations,
\begin{align}\label{eq_moments_SM}
\begin{split}
\langle \delta\theta_i^2 \rangle = \sigma^2
\sum_{\alpha,\beta}u_{\alpha,i}u_{\beta,i}u_{\alpha,i_0}u_{\beta,i_0}\frac{\tau_0 \left[\tau_0^3 \left(2 {\gamma}^2 m ({\lambda_\alpha}+{\lambda_\beta})+({\lambda_\alpha}-{\lambda_\beta})^2\right)+2 {\gamma} m \tau_0^2 \left(2 {\gamma}^2 m+{\lambda_\alpha}+{\lambda_\beta}\right)+8 {\gamma}^2 m^2 \tau_0+4 {\gamma} m^2\right]}{\left(2 {\gamma}^2 m ({\lambda_\alpha}+{\lambda_\beta})+({\lambda_\alpha}-{\lambda_\beta})^2\right) \left({\gamma} m \tau_0+{\lambda_\alpha} \tau_0^2+m\right) \left({\gamma} m \tau_0+{\lambda_\beta} \tau_0^2+m\right)} \, ,
\end{split}
\end{align}
The validity of Eq.~\eqref{eq_moments_SM} is confirmed numerically in Fig.~\ref{figS1}, for a general case where the noise correlation is neither long, nor short,
$\lambda_2 < \tau_0^{-1} < \lambda_N$.

In the long correlation time limit, it can furthermore be shown that the $p^{\rm th}$ cumulant is given by the $p^{\rm th}$ power of the Green's function.
Higher moments can also be calculated in the finite and
short correlation time limit, however the expressions become rather long and do not bring much direct information.
We here only discuss
the skewness in the short correlation time limit. It reads
\begin{eqnarray}
\begin{split}\label{eq:skewness}
\langle \delta\theta_j^3 \rangle = a_3 \,\sigma^3\, \tau_0^2 \sum_{\alpha,\beta,\gamma\ge 2} & u_{\alpha,i_0}u_{\beta,i_0}u_{\gamma,i_0}u_{\alpha,j}u_{\beta,j}u_{\gamma,j} \\
& \times \frac{{8 md^2 (\lambda_\alpha+\lambda_\beta+\lambda_\gamma )}+{24 d^4}-2m^2 \left(\lambda_\alpha^2 + \lambda_\beta^2 + \lambda_\gamma^2 -2 \lambda_\alpha\lambda_\beta -2\lambda_\alpha\lambda_\gamma -2\lambda_\beta\lambda_\gamma\right)}{A+B+C+D} \, ,
\end{split}
\end{eqnarray}
with
\begin{eqnarray*}
A&=&{4 md^4 \left(7 \lambda_\alpha^2+7 \lambda_\beta^2 +7 \lambda_\gamma ^2+2 \lambda_\alpha\lambda_\beta+2 \lambda_\beta\lambda_\gamma +2 \lambda_\alpha \lambda_\gamma \right)} \, , \\
B&=&2 m^2d^2 \left(-5 \lambda_\alpha^2 \lambda_\beta-5 \lambda_\alpha^2 \lambda_\gamma +5 \lambda_\alpha^3-5 \lambda_\alpha \lambda_\beta^2+42 \lambda_\alpha \lambda_\beta \lambda_\gamma -5 \lambda_\alpha \lambda_\gamma ^2-5 \lambda_\beta^2 \lambda_\gamma +5 \lambda_\beta^3-5 \lambda_\beta \lambda_\gamma ^2+5 \lambda_\gamma ^3\right) \, , \\
C&=&{24 d^6 (\lambda_\alpha+\lambda_\beta+\lambda_\gamma )} \, ,\\
D&=&m^3 \left(\lambda_\alpha^2 + \lambda_\beta^2 + \lambda_\gamma^2 -2 \lambda_\alpha\lambda_\beta -2\lambda_\alpha\lambda_\gamma -2\lambda_\beta\lambda_\gamma\right)^2 \, .
\end{eqnarray*}
We easily check that, in the $m=0$ limit, we recover the result for Kuramoto oscillators,
\begin{eqnarray}
\langle \delta\theta_j^3 \rangle &=& a_3 \, \sigma^3\, \tau_0^2\sum_{\alpha,\beta,\gamma\ge 2}\frac{u_{\alpha,i_0}u_{\beta,i_0}u_{\gamma,i_0}u_{\alpha,j}u_{\beta,j}u_{\gamma,j}}{d^2 (\lambda_\alpha+\lambda_\beta+\lambda_\gamma )} \, .
\end{eqnarray}

\begin{figure}[h]
\includegraphics[width=0.6\textwidth]{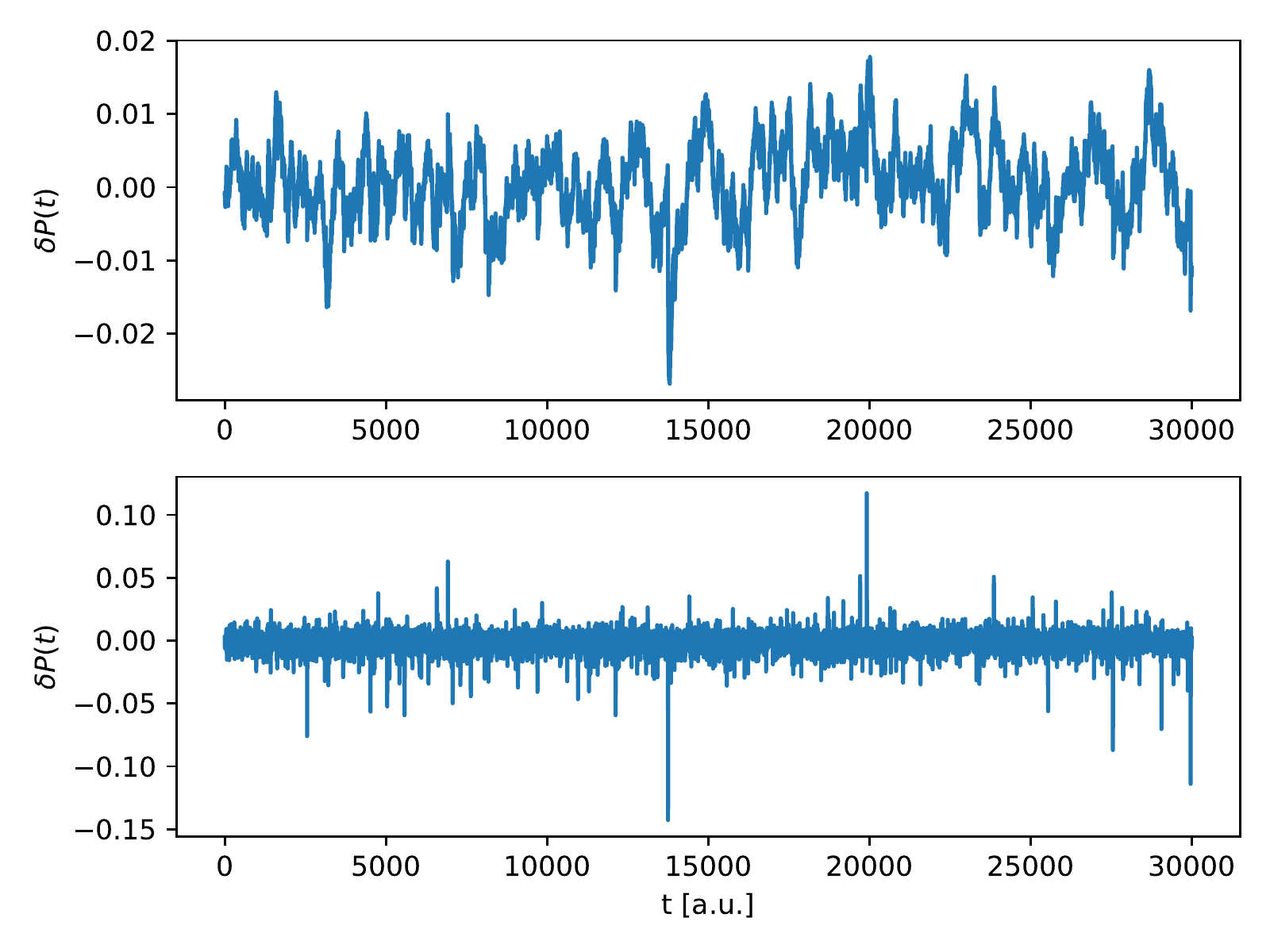}
\caption{Examples of noise sequences generated using Eq.~(\ref{eq:noise}) for two different correlation times $\tau_0$\,.}\label{figS2}
\end{figure}
\section{Numerical Simulations and Noise Generation}
All numerical simulations are done by time-evolving Eq.~(1)
 in the main text using standard fourth-order Runge-Kutta algorithms. More interesting is the generation of time-correlated non-Gaussian noise. To do that we follow the standard procedure to generate time-correlated noise as an Ornstein-Uhlenbeck process,
 \begin{eqnarray}\label{eq:noise}
\delta   \dot{P}(t)= -\tau_0^{-1}\delta P(t) + \delta P_0\, \xi(t)\,.
 \end{eqnarray}
However, instead of taking $\xi$ as a normal random variable, we take it as a non-Gaussian process. In the numerical simulations presented in the main text, we take it as a partially de-symmetrized log-normal process. This simple trick allows to generate noise sequences with finite correlation time and non-Gaussian properties. Figure~\ref{figS2} shows a few example of noise sequences generated using Eq.~(\ref{eq:noise}). The top panel has a longer correlation time than the bottom one.

\section{Correlation Time in New Renewable Power Feed-in Noise}

\begin{figure}[h]
\includegraphics[width=0.24\textwidth]{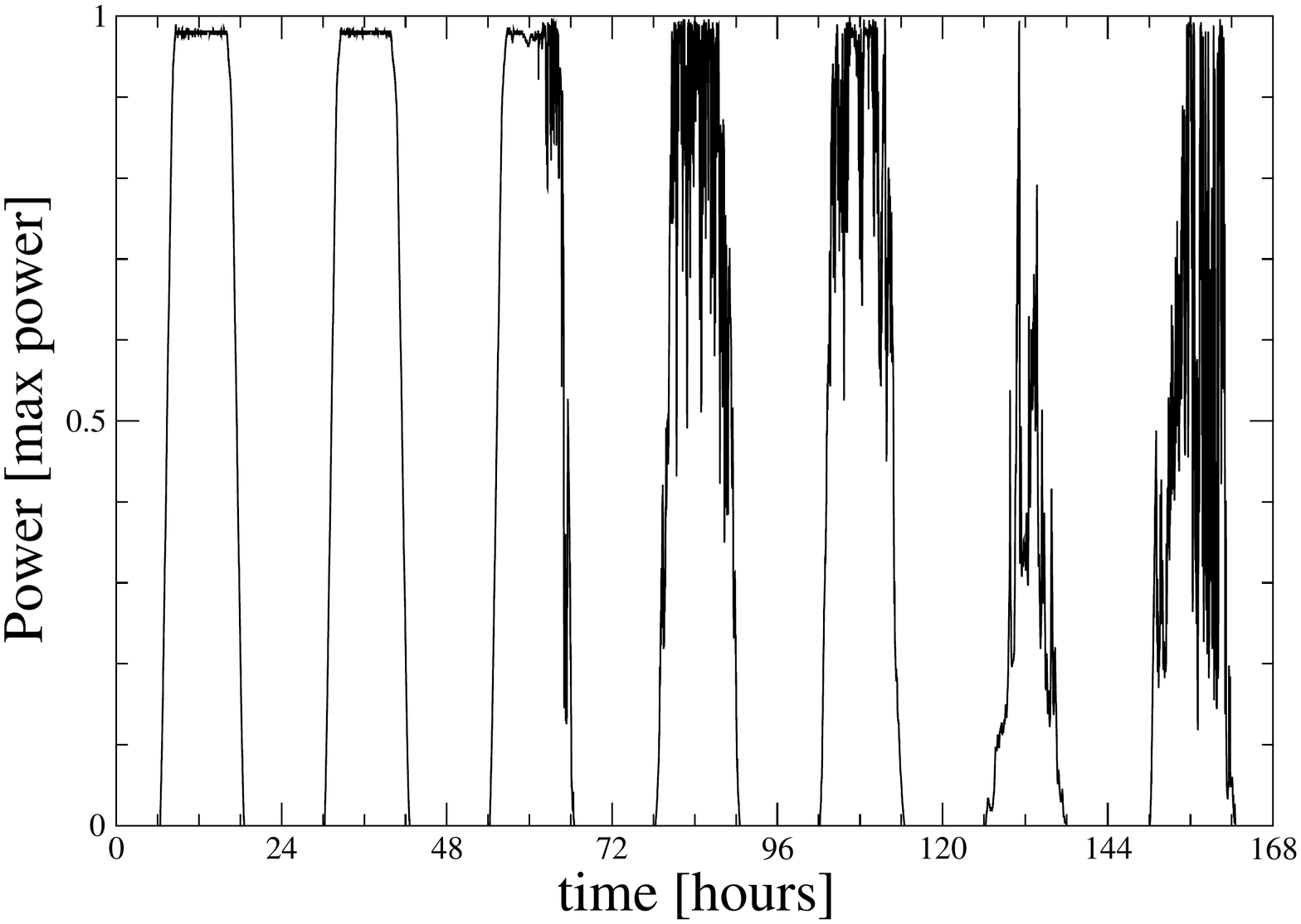}
\includegraphics[width=0.24\textwidth]{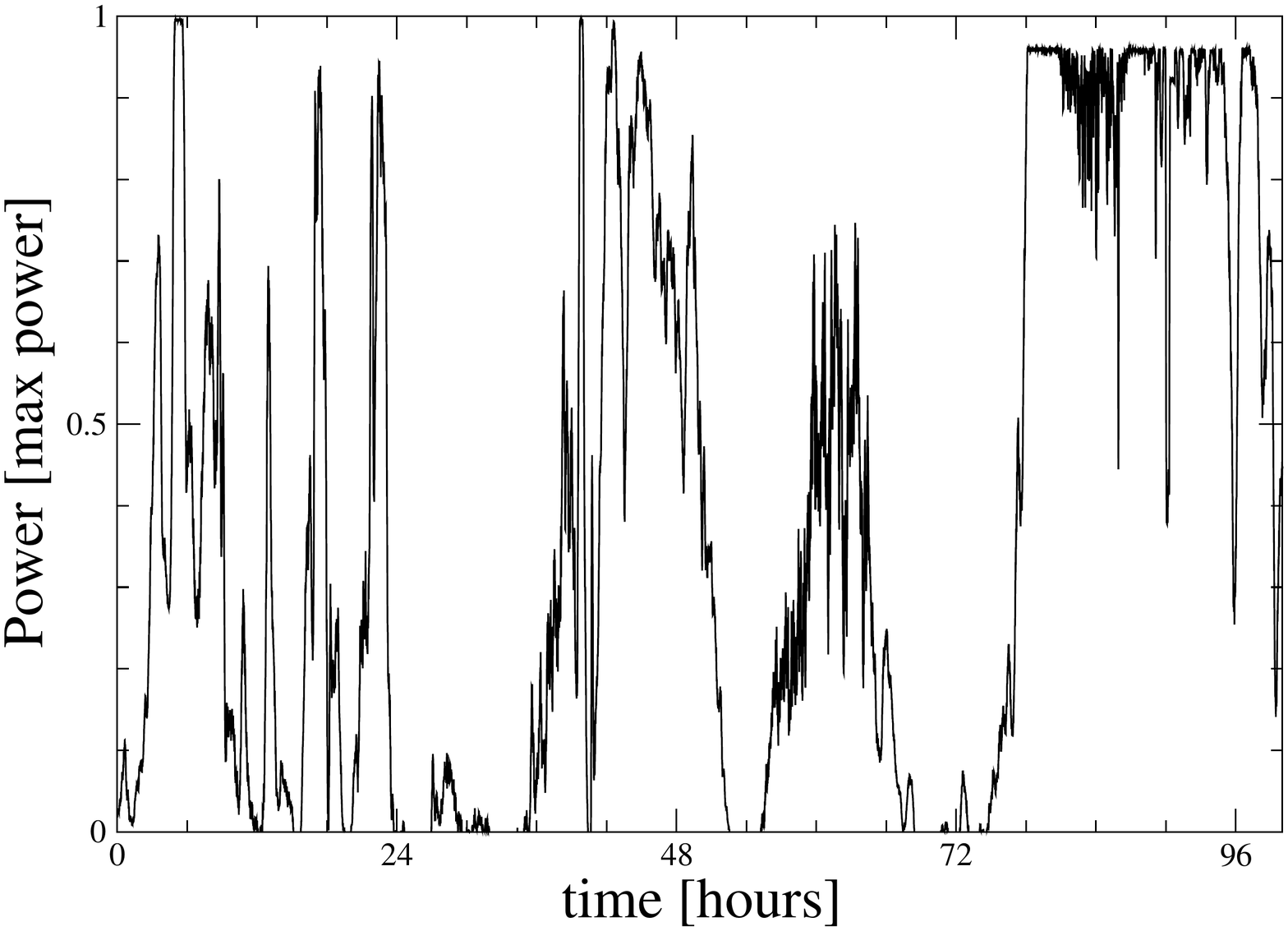}
\includegraphics[width=0.24\textwidth]{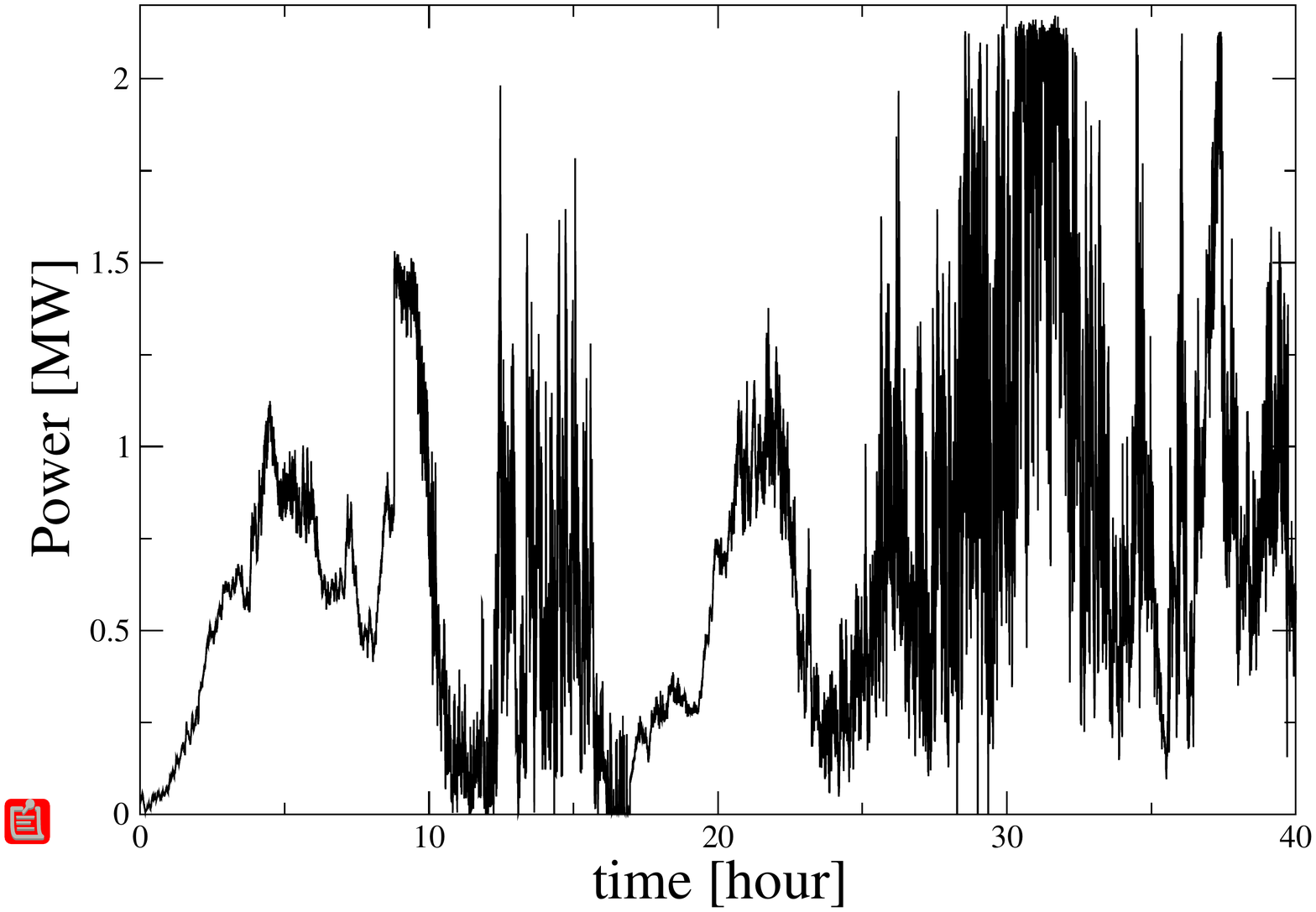}
\includegraphics[width=0.24\textwidth]{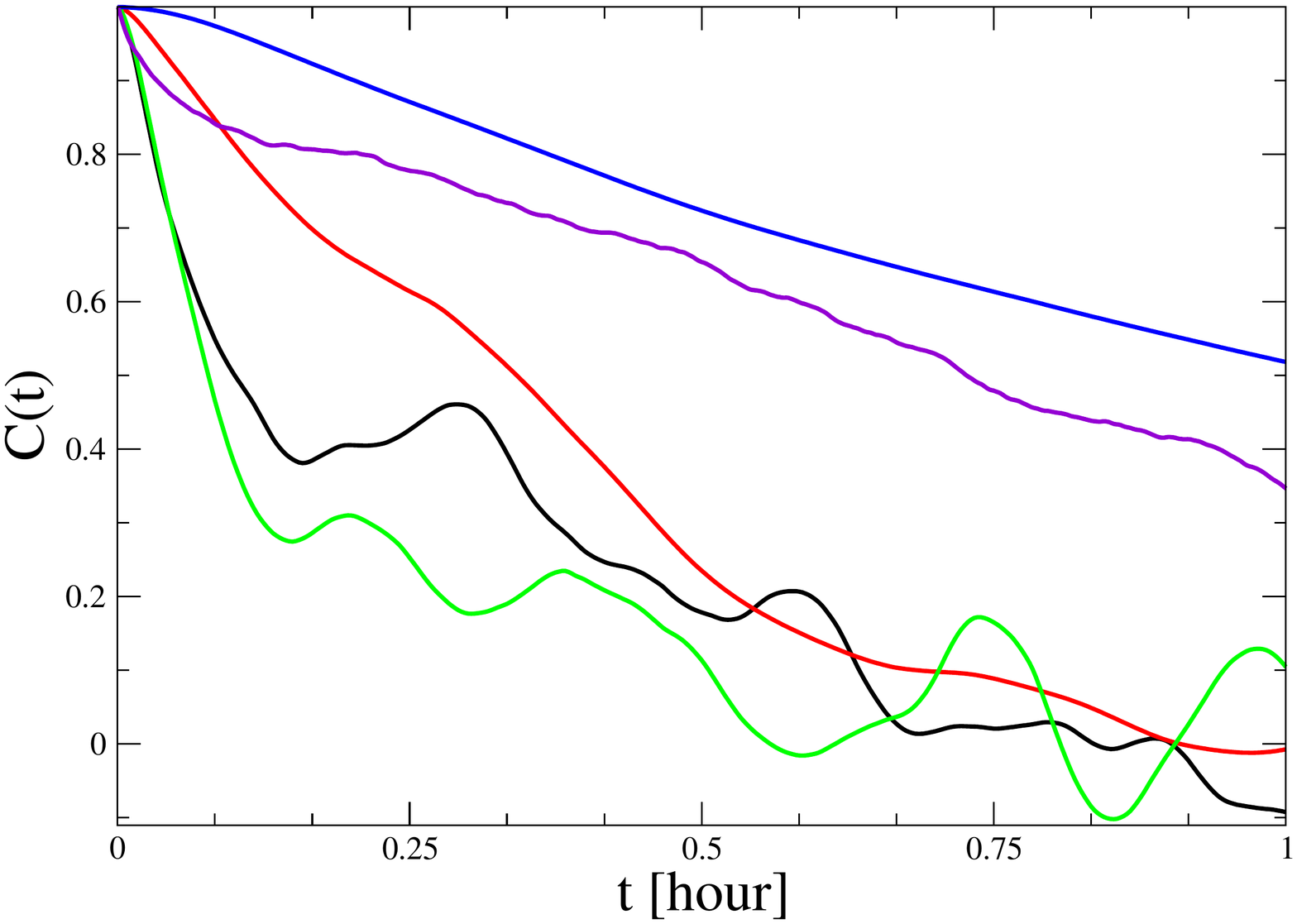}
\caption{Typical power feed-in fluctuations for a photovoltaic panel farm in Arizona (left; power rescaled with maximal power), a wind turbine farm in Arizona
(middle left; power rescaled with maximal power) and a 2 MW wind turbine in Germany (middle right).
Data acquisition frequencies are 0.1 Hz (left and middle left) and 1Hz (middle right). Power production correlator $C(t)$ (see text) for the fifth (black line) sixth (red) and seventh (green) PV production days
shown in the leftmost panel, and for the Arizona wind turbine farm (blue) and the Germany wind turbine (violet).
Data courtesy of Will Holgmren/Tucson Electric Power (left and middle left) and Ref.~\cite{data} (middle right).}\label{figS3}
\end{figure}

In Fig.~\ref{figS3} we show power feed-in time series for photovoltaic and wind turbine productions. Fluctuations magnitude and correlation time depend on production type, geographical location and time, i.e. seasonal and daily weather fluctuations.
Here, we are interested in getting qualitative information on typical correlation time. From the data of Fig.~\ref{figS3} we extract the normalized correlator $C(t)=\langle \delta P_i(t_0)  \delta P_i(t_0+t)\rangle/ \langle \delta P_i^2(t_0) \rangle$.
For the photovoltaic data, the correlator is calculated over production periods of 6 hours, individually for each of the seven days shown in the left panel of Fig.~\ref{figS3}, while for the wind turbines, the correlator is calculated for the full duration shown in
Fig.~\ref{figS3}.

The rightmost panel of Fig.~\ref{figS3} shows the correlators for the three most strongly fluctuating days of PV production and for the full wind turbine production sequences. It is seen that the correlator decays over a scale of 1-2 minutes or more.
Given that intrinsic time scales of large-scale power grids lie in the subsecond range, we conclude that new renewable power productions are in the long correlation time regime.

\end{document}